\documentclass[sigconf]{acmart}
\AtBeginDocument{%
  \providecommand\BibTeX{{%
    \normalfont B\kern-0.5em{\scshape i\kern-0.25em b}\kern-0.8em\TeX}}}

\setcopyright{rightsretained}
\copyrightyear{2023}
\acmYear{2023}
\acmDOI{10.1145/3544548.3580670}

\acmConference[CHI '23]{Proceedings of the 2023 CHI Conference on Human Factors in Computing Systems}{April 23--28, 2023}{Hamburg, Germany}
\acmBooktitle{Proceedings of the 2023 CHI Conference on Human Factors in Computing Systems (CHI '23), April 23--28, 2023, Hamburg, Germany}
\acmISBN{978-1-4503-9421-5/23/04}

\usepackage{multirow}
\usepackage[]{natbib}%
\usepackage{xr-hyper,refcount}
\usepackage{graphicx}
\usepackage[utf8]{inputenc} %
\usepackage{url}            %
\usepackage{booktabs}       %
\usepackage{amsfonts}       %
\usepackage{microtype}      %
\usepackage{algorithm,algorithmic}
\usepackage{amsmath,amsthm}
\usepackage{mathtools}
\usepackage{subcaption}
\usepackage{xcolor}
\usepackage{color}
\usepackage{appendix}
\usepackage{xspace}
\usepackage{enumitem}
\usepackage{mathtools}
\usepackage{enumitem}
\usepackage{booktabs}
\usepackage[]{xcolor}
\usepackage{verbatim}

\newcommand{\pr}[1]{\mathcal{P}^{\text{#1}}}

\newcommand{\cycle}{\text{CYCLE}}
\newcommand{\repeated}{\text{REPEAT}}

\newcommand{\E}{\mathbb{E}}
\definecolor{OliveGreen}{rgb}{0,0.6,0}

\usepackage{mathtools}

\begin{document}

\title[Understanding the Validity of Assumptions on Human Preferences in Multi-armed Bandits]{A Field Test of Bandit Algorithms for Recommendations: Understanding the Validity of Assumptions on Human Preferences in Multi-armed Bandits}

\author{Liu Leqi}
\authornote{Both authors contributed equally to this research.}
\affiliation{%
  \institution{Carnegie Mellon University}
    \city{Pittsburgh}
  \country{USA}
}
\email{leqil@cs.cmu.edu}

\author{Giulio Zhou}
\authornotemark[1]
\affiliation{%
  \institution{Carnegie Mellon University}
    \city{Pittsburgh}
  \country{USA}
  }
\orcid{0000-0002-9802-0741}
\email{gzz@cs.cmu.edu}

\author{Fatma K{\i}l{\i}n\c{c}-Karzan}
\affiliation{%
  \institution{Carnegie Mellon University}
    \city{Pittsburgh}
  \country{USA}
  }
\email{fkilinc@andrew.cmu.edu}

\author{Zachary C. Lipton}
\affiliation{%
  \institution{Carnegie Mellon University}
  \city{Pittsburgh}
  \country{USA}
  }
\email{zlipton@andrew.cmu.edu}

\author{Alan L. Montgomery}
\affiliation{%
  \institution{Carnegie Mellon University}
 \city{Pittsburgh}
  \country{USA}
  }
\email{alm3@andrew.cmu.edu}

\begin{abstract}
Personalized recommender systems suffuse modern life, 
shaping what media we read and what products we consume.
Algorithms powering such systems tend to consist 
of supervised-learning-based heuristics,
such as latent factor models
with a variety of heuristically chosen prediction targets.
Meanwhile, theoretical treatments of recommendation 
frequently address the decision-theoretic nature of the problem,
including the need to balance exploration and exploitation,
via the multi-armed bandits (MABs) framework.
However, MAB-based approaches rely heavily on assumptions 
about human preferences. These preference assumptions are seldom tested using human subject studies, partly due to the lack of publicly available toolkits to conduct such studies.
In this work, we conduct a study with crowdworkers in a comics recommendation MABs setting. Each arm represents a comic category, and users provide feedback after each recommendation. We check the validity of core MABs assumptions---that human preferences (reward distributions) are fixed over time---and find that they do not hold. This finding suggests that any MAB algorithm used for recommender systems should account for human preference dynamics. While answering these questions, we provide a flexible experimental framework for understanding human preference dynamics and testing MABs algorithms with human users. The code for our experimental framework and the collected data can be found at \href{https://github.com/HumainLab/human-bandit-evaluation}{https://github.com/HumainLab/human-bandit-evaluation}. 
\end{abstract}



\maketitle

\section{Introduction}\label{sec:intro}

As online services have become inescapable fixtures of modern life,
recommender systems have become ubiquitous,
influencing the music curated into our playlists,
the movies pumped into the carousels of streaming services,
the news that we read,
and the products suggested whenever we visit e-commerce sites.
These systems are commonly data-driven and algorithmic, 
built upon the intuition that historical interactions 
might be informative of users' preferences,
and thus could be leveraged
to make better recommendations~\citep{gomez2015netflix}.
While these systems are prevalent in real-world applications, 
we often observe misalignment between their behavior
and human preferences~\citep{jain2015trends}. 
In many cases, such divergence comes from the fact 
that the underlying assumptions powering 
the learning algorithms are questionable. 

Recommendation algorithms for these systems 
mostly rely on supervised learning heuristics~\citep{bennett2007netflix,gomez2015netflix},
including latent factor models
such as matrix factorization~\citep{koren2009matrix}
and deep learning approaches
that are designed to predict various heuristically chosen targets~\citep{wei2017collaborative}
(e.g., purchases, ratings, reviews, watch time, or clicks~\citep{bennett2007netflix, mcauley2013amateurs}).
Typically they rely on the naive assumption
that these behavioral signals 
straightforwardly indicate the users' preferences.
However, this assumption may not hold true in general
for a variety of reasons, including exposure bias
(users are only able to provide behavioral signals
for items they have been recommended)
and censoring 
(e.g., reviews tend to be written 
by users with strong opinions)~\citep{swaminathan2015counterfactual, joachims2017unbiased}.

On the other hand, 
to study the decision-theoretic nature of recommender systems,   
the online decision-making framework---multi-armed bandits (MABs)---has commonly been used~\citep{slivkins2019introduction,lattimore2020bandit}.
In MABs, at any given time, 
the decision-maker chooses an arm to pull 
(a recommendation to make in the recommender system setting) 
among a set of arms 
and receives a reward,
with the goal of 
obtaining high expected cumulative reward over time.
Theoretical research on MABs 
centers on algorithms that 
balance between 
the exploration and exploitation tradeoff 
and analyses capturing the performance\footnote{We provide more details on how performances are defined in MABs in Section~\ref{sec:mab}.} of these algorithms \citep{lattimore2020bandit}.
There is a long line of work on developing MAB-based approaches for recommender systems in settings including traditional $K$-armed bandits  \citep[and references therein]{barraza2020introduction}, contextual bandits \citep[and references therein]{li2010contextual,mcinerney2018explore,mehrotra2020bandit}, and Markov decision processes~\citep[and references therein]{shani2005mdp,chen2019top,chen2021user,chen2021values,wang2022surrogate,chen2022off}.
To guide {such research} on applying MAB-based algorithms in recommender systems, %
it is of importance to \emph{test} 
whether the assumptions that these algorithms are built upon 
are valid in real-world recommendation settings.

In this work, we focus on the assumption
of temporal stability that 
underlies both practical supervised learning methods
and algorithms for classical MAB settings
where the reward distribution 
is assumed to be fixed over time.
In a recommendation setting,
the reward distribution of an arm
corresponds to the user's preference towards 
that recommendation item. 
Although the assumption that the reward distribution is fixed
may be appropriate to applications driving early MABs research 
(e.g., sequential experimental design in medical domains)~\citep{robbins1952some},  
one may find it to be unreasonable in recommender systems 
given that the interactants are humans 
and the reward distributions represent human preferences. 
For example, consider the task of restaurant recommendations, 
though a user may be happy
with a recommended restaurant for the first time, 
such enjoyment may decline as the same recommendation 
is made over and over again.  
This particular form of evolving preference  is known as {satiation}, 
and results from repeated consumption~\citep{galak2018properties}.
One may also think of cases 
where a user's enjoyment increases
as the same item being recommended multiple times, 
due to reasons including sensitization~\citep{groves1970habituation}. %
In both settings, the assumption that reward distributions are fixed is violated
and the recommendation algorithms 
may influence the preferences of their users.

We test the assumption on fixed reward distributions 
through randomized controlled trials
conducted on Amazon Mechanical Turk. 
In the experiment, we simulate a $K$-armed bandit setting 
(a MAB setup where the arm set is the same set of $K$ arms over time)
and recommend comics from $K$ comic series to the study participants. 
After reading a comic, 
the participants provide an enjoyment score 
on a $9$-point Likert scale~\citep{cox1980optimal}, 
which serves as the reward received 
by the algorithm for the recommendation 
(for pulling the corresponding arm). 
Each comic series belongs to a different genre 
and represents an arm. 
Our analyses on the collected dataset
reveal that in a bandit recommendation setup, 
human preferences can evolve, 
even within a short period of time 
(less than $30$ minutes) 
(Section~\ref{sec:evolving-preference}). 
In particular, 
between two predefined sequences 
that result in the same number of pulls
of each arm in different order, 
the mean reward  for one arm 
has a significant difference of $0.57$ 
($95\%$ CI = $[0.30, 0.84]$, $p$-value $< 0.001$).
This suggests that any MAB algorithms
that are applied to recommendation settings
should account for 
the dynamical aspects of human preferences
and the fact that 
the  recommendations 
made by these algorithms may influence users' preferences.

The line of work that develops contextual bandits and reinforcement learning algorithms for recommender systems \citep{shani2005mdp, mcinerney2018explore, chen2019top, mehrotra2020bandit, chen2021user, chen2021values, wang2022surrogate, chen2022off} hinges upon the assumption that users' preferences depend on the past recommendations they receive. The proposed algorithms have been deployed to interact with users on large-scale industry platforms (e.g., Youtube). 
These prior works differ from ours in multiple ways.
Among them, the most significant distinction is that 
our work aims to understand and identify assumptions on reward distributions that better represent user preference characteristics.
On the other hand, this line of work asks the question that \emph{once an assumption on the reward distribution has been made}, how to design the recommendation algorithms so that they have better performance.
For example, in settings where recommender systems are modeled as contextual bandits~\citep{mehrotra2020bandit, mcinerney2018explore},
the reward distributions are assumed to take a certain functional form (often linear) in terms of the observable states.
When treating recommender systems as reinforcement learning agents in a Markov decision process~\citep{shani2005mdp, chen2019top, chen2021user, chen2021values, wang2022surrogate, chen2022off}, 
one has made the core assumption that the reward is Markovian and depends only on the \emph{observable} user states (e.g., user history) and recommendations. 
In other words, the reward (and user preference) does not depend on \emph{unobservable} states (e.g., user emotions) as in partially observable Markov decision processes.
Under this assumption, the proposed algorithms deal with  difficulties (e.g., large action spaces) in the reinforcement learning problem. 
It is worth noting that in these prior works, the proposed algorithms have been evaluated on industry platforms.
For academics who want to evaluate their recommendation algorithms with human interactants, such infrastructure is not easily accessible. We take an initial step to address this need by developing our experimental framework.

The experimental framework we developed is flexible.
It allows one to conduct field tests of MAB algorithms,
use pre-defined recommendation sequences to analyze human preferences,
and ask users to choose an arm to pull on their own. 
These functionalities can be used to identify assumptions on user preference dynamics and reward distributions that better capture user characteristics.
As an illustration of the flexibility of our experimental framework, 
we have collected data while the participants interact
with some traditional MAB algorithms
and analyze their experience with these algorithms.
Interestingly, we observe that interactants
(of a particular algorithm) 
who have experienced the lowest level of satisfaction
are the ones to have the poorest performance in recalling 
their past ratings for previously seen items.

In summary, %
we provide a flexible experimental framework
that can be used to run field tests with humans
for any $K$-armed bandit algorithms (Section~\ref{sec:framework}). 
Using this experimental framework, we have tested the validity of 
the fixed-reward-distribution (fixed-user-preference) assumption 
for applying MAB algorithms to recommendation settings (Section~\ref{sec:evolving-preference}). 
As an illustration of the flexibility of our experimental framework, 
we have {inspected different bandit algorithms in terms of user enjoyment and attentiveness} (Section~\ref{sec:diff-algo}). 
We discuss the limitation of our study in Section~\ref{sec:discussion}.

\begin{figure*}
    \centering
    \includegraphics[width=\textwidth]{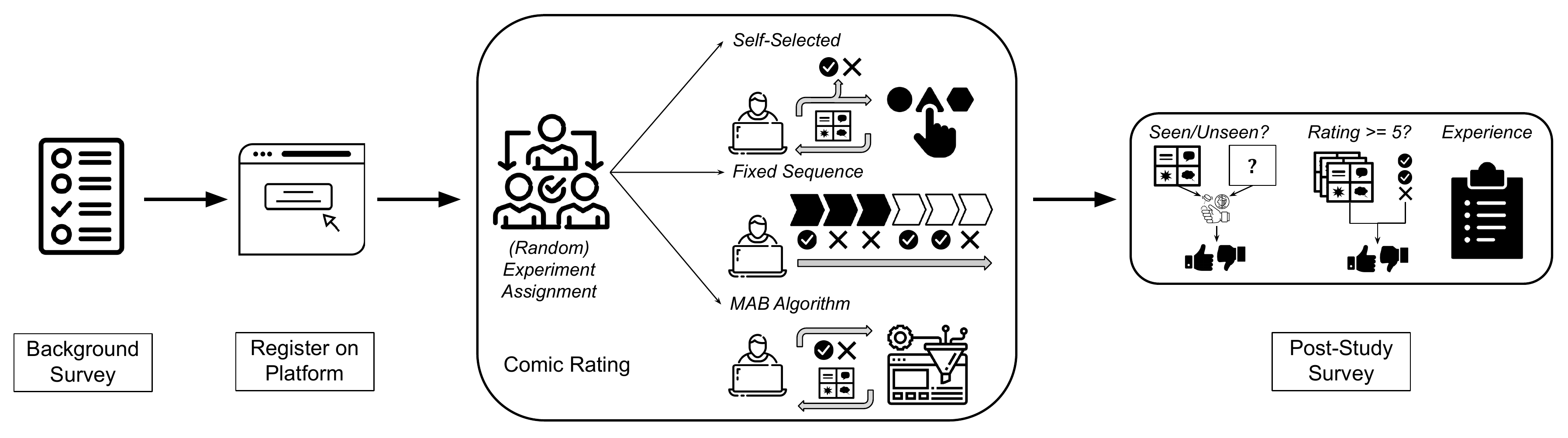}
    \caption{Overview of the experimental protocol. Participants first complete a background survey and then register their MTurk ID on our platform to get randomly assigned (without their direct knowledge) one of the following types of experimental setups: Self-selected, one of the fixed sequences, or one of the MAB algorithms. Study participants who are assigned to a fixed sequence and an MAB algorithm  only provide ratings (enjoyment scores to the comics). %
    Self-selected study participants provide both a rating as well as the next genre to view. Once the comic rating portion of the study is complete, participants move onto the post-study survey, where they are asked questions related to their experience in the study, e.g., how well they remember the consumed content. Participants must complete all parts of the study and answer attention checks sufficiently correctly to receive full compensation, and therefore be included in the final study data. }  %
    \label{fig:workflow}
\end{figure*}

\section{Related Work}\label{sec:related-work}

The study of evolving preferences has a long history,
addressed by such diverse areas as psychology~\citep{christensen2006changing, galak2018properties}, 
economics~\citep{prelec2004decreasing,gul2005revealed}, marketing~\citep{tucker1964development},  operations research~\citep{baucells2007satiation},  philosophy~\citep{liu2007changing,baber2007adaptive},
and recommender systems~\citep{kapoor2015just,rafailidis2015modeling, lee2014modeling}.
In the bandits literature, 
there is a recent line of work, 
motivated by recommender systems,
that aims to incorporate the dynamic nature of human preference into the design of algorithms. 
These papers have different models 
on human preferences, 
expressed as specific forms 
of how the reward of an arm depends on 
the arm's past pulls.
\citet{levine2017rotting, seznec2018rotting} 
model rewards as monotonic functions
of the number of pulls.
By contrast,
\citet{kleinberg2018recharging, basu2019blocking, cella2020stochastic} 
consider the 
reward to be a 
function of the time elapsed
since the last pull of the corresponding arm. 
In \citet{mintz2020nonstationary}, 
rewards  are context-dependent, 
where the contexts are updated based on known deterministic dynamics.
Finally, \citet{leqi2020rebounding}
consider the reward dynamics to be unknown
stochastic linear dynamical systems. 
These prior works model the reward (user preferences) in distinct ways,
and lack (i) empirical evidence 
on whether user preferences 
 evolve in the short period of time 
in a bandit setup;
and (ii) datasets and experimental toolkits that can be used to 
verify the proposed theoretical models
and to explore more realistic ways of modeling user preferences.
On the other hand, there is a line of work on developing contextual bandits and reinforcement learning algorithms to account for user preference dynamics in recommender systems. The evaluations of  these algorithms against human users rely on accessibility to large-scale industry platforms~\citep{shani2005mdp, mcinerney2018explore, chen2019top, mehrotra2020bandit, chen2021user, chen2021values, wang2022surrogate, chen2022off}.

Datasets that are publicly available and can be used to 
evaluate bandit algorithms in recommendation settings 
often contain ratings or other types of user feedback 
of recommendations~\citep{saito2020open,lefortier2016large}. 
These datasets do not contain 
trajectories of recommendations 
and the associated feedback signal for a particular user, 
making it hard to understand the dynamical aspects
of user preferences and to identify effects
of past recommendations on the preferences. 
In addition, 
existing MAB libraries (e.g.,~\citep{BanditPyLib})
only consist of implementations of bandit algorithms,
but lack the appropriate tools and infrastructure 
to conduct field tests of these algorithms 
when interacting with humans.
The toolkit we have developed closes this gap
and allows one to use these libraries 
while conducting human experiments 
on bandit algorithms. 
Another relevant stream of research that 
considers human experiments in bandits settings
are the ones that 
ask human participants to make decisions
in a bandit task 
and collect data for modeling their decision-making  ~\citep{acuna2008bayesian,reverdy2014modeling,lee2011psychological}. 
In other words, in those experiments,
human participants take the role of the algorithm that selects the arm to pull
instead of the role of providing reward signals. 
In one of our experimental setups, 
the study participants are asked to 
select comics to read on their own. 
However, in contrast to reward distributions 
that are defined by the experiment designers of those experiments,
in our setting, 
the rewards are provided by the human participants, 
indicating their preferences.

\section{Experimental Setup}
\label{sec:experimental-setup}

\begin{figure*}
    \centering
    \includegraphics[width=.8\textwidth]{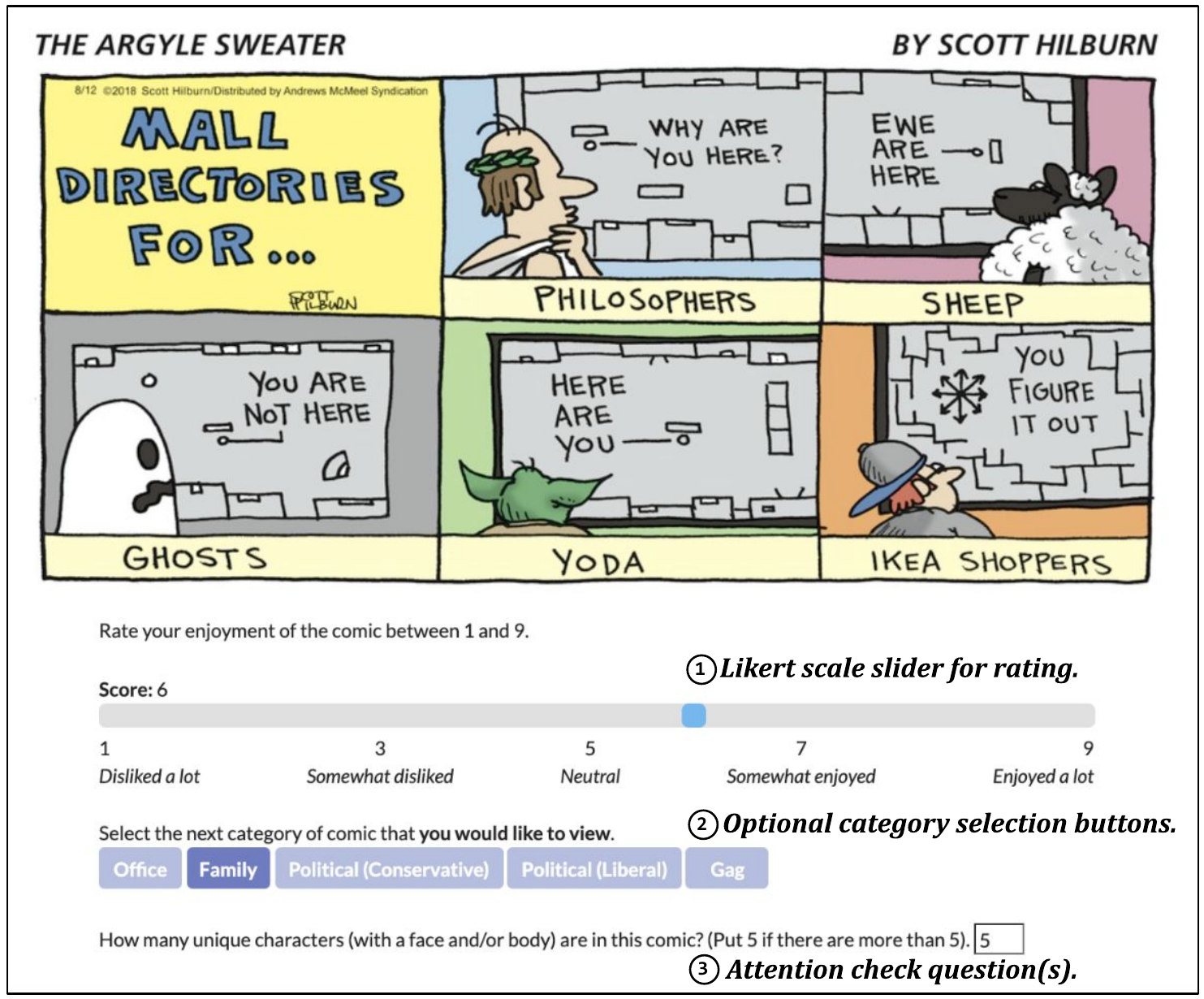}
    \caption{The user interface of our experimental platform when the participants read and rate a comic. 
    For each comic, the participants have up to three action items to complete. First, they must provide the comic an enjoyment score (a rating) between $1$ and $9$ using the Likert scale slider bar, indicating how they like the comic. Second, if the participants are under the Self-selected setting, 
    they must select the genre of comic they would like to view next. 
    For other participants, this step does not exist. 
    Finally, the participants are asked to answer one or more customized attention check questions.
    } \label{fig:ui}
\end{figure*}

In this section, we first describe 
a classical MABs setting---stochastic $K$-armed bandits---and
discuss the algorithms we have used in our experiments (Section~\ref{sec:mab}). 
We then provide reasoning on why we choose comics as the reommendation domain
and selection criteria
for the comics used for recommendations (Section~\ref{sec:dataset}). 
Finally, we discuss 
our experimental framework (Section~\ref{sec:framework}).

\subsection{Stochastic $K$-armed bandits}\label{sec:mab}
In stochastic $K$-armed bandits, 
at any time $t$, 
the learner (the recommender in our case)
chooses an arm $a(t) \in [K]$ to pull (a recommendation to present in our case)
and receives a reward $R(t)$. 
For any horizon $T$,
the goal for the learner is to attain 
the highest expected cumulative reward $\E[\sum_{t=1}^T R(t)]$.
In this setup, the reward distribution of each arm $k \in [K]$
is assumed to be fixed over time
and the rewards received by pulling the same arm 
are independently and identically distributed. 
An oracle (the best policy), in this case, 
would always play the arm 
with the highest mean reward.
The difference between the expected cumulative reward 
 obtained by the oracle
and the one  obtained by the learner
is known to be the ``regret.''
As is shown in existing literature~\citep{lattimore2020bandit},
the regret lower bound for stochastic $K$-armed bandits 
is $\Omega(\sqrt{T})$.
Many existing algorithms 
achieve the regret at the optimal rate $O(\sqrt{T})$, 
including the Upper Confidence Bound (UCB) algorithm 
and the Thompson Sampling (TS) algorithm.
We also include a traditional algorithm Explore-then-Commit (ETC)
and a greedy heuristic ($\varepsilon$-Greedy)
in our study. 
These algorithms along with their regret guarantees
are built upon the key assumption that the reward distributions are fixed. 
In Section~\ref{sec:evolving-preference},
we test the validity of this assumption in a recommendation setting 
where the interactants are humans 
and the rewards represent their preferences.

\paragraph{Algorithms}
We give a brief summary and a high-level intuition for each algorithm. 
More detailed descriptions of these algorithms
can be found in Appendix~\ref{appendix:algorithms} in the supplementary material.
We use the term ``algorithm''
in a broad sense 
and the following items (e.g., Self-selected and Fixed sequence) may not all be traditional MAB algorithms.

\begin{itemize}
    \item Self-selected: The participants who are assigned to the Self-selected algorithm will choose which arm to interact with by themselves. 
    In other words, at each time $t$, instead of a prescribed learning policy determining the arm $a(t)$, the participants themselves will choose the arm.  
    \item UCB: %
    At time $t$, UCB deterministically pulls the arm with the highest upper confidence bound, a value that for each arm, combines the empirical mean reward with an uncertainty estimate of the mean reward. 
    An arm with high upper confidence bound can have high empirical mean and/or high uncertainty on the mean estimate. 
    \item TS: In Thompson Sampling, a belief distribution is maintained over the possible reward values for each arm. 
    At time $t$, the algorithm samples a reward from each arm's belief distribution 
    and pulls the arm with the highest sampled reward.
    When a reward is received after pulling the arm, TS updates the corresponding arm's prior belief distribution to obtain a posterior. 
    \item ETC: 
    Unlike UCB and TS, the Explore-then-Commit algorithm has two separate stages---the exploration stage and the exploitation stage. 
    It starts with an exploration period where the algorithm pulls the arms in a cyclic order
    and then switches to an exploitation stage where only the arm with the highest empirical mean in the exploration stage will be pulled. 
    Given that the ETC algorithm achieves a regret of $O(T^{2/3})$ when the exploration time is on the order of $T^{2/3}$,
    we have set the exploration period to be $c \cdot T^{2/3}$ %
    for a positive constant $c$. 
    \item $\varepsilon$-Greedy: 
    This greedy heuristic pulls the arm with the highest empirical mean with probability $1-\varepsilon$ where $0 < \varepsilon < 1$, 
    and pulls an arm uniformly at random with probability $\varepsilon$. 
    In a setting with long interaction period,
    one way of setting $\varepsilon$ is to 
    have it decreasing over time, e.g., setting $
    \varepsilon$ to be on the order of $\frac{1}{t}$~\cite{auer2002finite}.
    Given the short interaction period in our setting, 
    we have used a fixed $\varepsilon = 0.1$ (which may result in linear regret). 
    \item Fixed sequence (CYCLE, REPEAT): 
    The fixed sequence algorithms pull arms by following a predefined sequence. 
    That is, the arm pulled at time $t$ only depends on the current time step and does not rely on the received rewards so far. 
    We have used two fixed sequences CYCLE and REPEAT for testing whether the reward distributions (that represent participants' preferences) are fixed over time.
    We provide more details on these two fixed sequences in Section~\ref{sec:evolving-preference}.
\end{itemize}

Next, we present 
how we have selected the comics used for recommendations.

\subsection{Comics data}\label{sec:dataset}
In our experiment, 
we choose comics 
as our recommendation domain for the following reasons:
(i) Fast consumption time: 
Given the nature of our experiment 
where study participants are recommended 
a sequence of items to consume in a
limited amount of time, 
we require the time for {consuming} %
each of the recommendations to be short.
For example, recommending movie clips to watch may not be appropriate in our setting given that each clip may take a couple of minutes to finish.
(ii) No strong {pre-existing preferences}:
Another important feature of 
the chosen recommendation domain 
is that the majority of the study participants should have no strong preference on that subject prior to the experiment. 
For example, unlike comics, 
music is a subject that most people 
tend to {already} have strong preference{s} towards~\citep{schafer2010makes}.
In such cases, 
the effects of recommendations towards 
the participants' preferences may be  minimal.

We collected comics from $5$ 
comic series on GoComics~\citep{gocomics}.  
Each comic series belongs to a genre
and represents an arm for pulling. 
The $5$ comic series along with 
their genre are 
Lisa Benson (political, conservative),
Nick Anderson (political, liberal),
Baldo (family),
The Born Loser (office),
and The Argyle Sweater (gag).
The genres of these comics 
are assigned by GoComics. 
For each series, we take the following steps
to select the set of comics: %
\begin{enumerate}
    \item We first collect all comics 
    belonging to the comic series
    from the year 2018. We select this time period to be not too recent so that the results of the study are not heavily influenced by ongoing events. It is also not too distant so that the content is still relevant to all subjects.
    \item For each individual comic, we obtain its number of likes on GoComics. Then, we choose the top $60$ comics from each comic genre/series in terms of the number of likes. 
    This selection criteria is designed to ensure the quality of the chosen comics.
    \item Finally, we randomly assign an ordering to the comics. We keep this ordering fixed throughout the study such that if an arm is pulled (a genre is chosen) at its $j$-th time, 
    the presented comics will always be the same.
\end{enumerate}
The comics within the same comic series are independent, 
{in the sense that they can generally be read in any order, without requiring much context from previous comics from the same series.}
For these collected comics, 
we have labeled the number of unique characters in them,
and use it for the  
attention check questions  
to ensure that the study participants 
have read the comics.  
Although we have adopted the above selection criteria to ensure the quality of comics from the same comic series to be similar, there is heterogeneity among individual comics and thus may influence the interpretation of our results. We provide more discussion on this in Section~\ref{sec:discussion}.
In Section~\ref{sec:framework},
we discuss our experimental protocol 
and platform.

\subsection{Experimental protocol and platform}
\label{sec:framework}

We first outline our experimental protocol,
which consists of the following steps (Figure~\ref{fig:workflow}):
\begin{enumerate}
    \item \textit{Background survey (initial filtering)}: We ask the study participants to complete a brief background survey  (Appendix~\ref{appenidx:ui-attention} in the supplementary material). 
    The participants will only be given a code to continue to the next step if an arithmetic question is answered correctly. 
    The goal for the first step is to set up an initial filtering for participants. 
    \item \textit{Registration}: After completing the background survey,
    participants are then asked to register on our platform
    using their completion code.
    Each participant in our %
    study is assigned an  
    algorithm in the following fixed probabilities---$0.25$ for Self-selected,
    $0.125$ for UCB, 
    $0.125$ for TS, 
    $0.125$ for ETC, 
    $0.125$ for $\varepsilon$-Greedy 
    and $0.125$ for each of the two fixed sequences.
    \item \textit{Comic rating}: In this step, participants 
    will read a sequence of $50$ comics and  provide an enjoyment score (a rating) after reading each comic.
    The sequence of comics can be generated by any of the algorithms discussed in Section~\ref{sec:mab}.
    After reading each comic and providing a rating, 
    the participants are also asked to answer an attention check question. %
    \item \textit{Post-study survey}: Once the participants are finished with the comics rating step of the study, 
    they are asked to complete a post-study survey about their reading experience.
    They are asked if they remember reading certain comics
    and if they have rated them  positively, 
    as well as how they perceive the recommendations they are provided.
    An example of the post-study survey questions can be found in Figure~\ref{fig:post-study-survey} (Appendix~\ref{appendix:implementation-details} in the supplementary material).
\end{enumerate}

Our experimental platform is built 
as a toolkit for running field tests for any MAB algorithms with human users.
It consists of (a) the participant-facing web interface,
and (b) the server backend that stores and processes incoming data from the web interface. 
When designing the experimental protocol and platform, 
we consider the following questions:
\begin{enumerate}
    \item Given that we are asking users to give subjective feedback, how do we have more user responses that are reflective to the user's true preference?
    \item How do we design an experimental interface 
    that impose less bias to the users?
    \item Since our study requires users to complete a long (up to 30-minute) sequence of non-independent tasks, 
    how do we have the study to be less interrupted?
    \item How do we build the system flexible enough to conduct studies for  different MAB algorithms and test different assumptions of MAB setups, including ones that we do not cover in this work?
\end{enumerate}

For (1), 
we adopt a 9-point Likert scale 
so that the numbers are sufficiently distinguishable to the participants~\citep{cox1980optimal} 
and check whether users are paying sufficient attention during the study.
In particular,
we test the participants on objective properties of the comics they have read,
e.g. the number of unique characters (with a face and/or body) in them.
We also set a minimum time threshold of 10 seconds before each user response can be submitted
so that users spend adequate time on each comic.

For (2),
to ensure that the participant's rating
is not biased towards (e.g., anchored on) the 
Likert scale slider's initial value, 
we set the slider to be transparent before the
participant clicks on the scale.
In addition, in the Self-selected setting
where the participants choose the genre
of comics to read next, 
we minimize the color- and ordering-based biases by
setting the category selection buttons to be the same color 
and in random order.

As stated in design question (3), 
because we are interested in studying evolving preferences over a sequence of non-independent tasks,
we would like to have \textit{continuous and uninterrupted} user attention over a period of time.
To do so,
we design the system to be stateful
so that participants can resume the study where they left off
in the event of brief network disconnection and browser issues.

Finally, we discuss the flexibility of our system
and address design question (4). 
Our experimental platform 
allows the experimenter to specify 
any recommendation domains 
and MAB algorithms for interacting with the human interactant. 
This flexibility allows the experimenter 
to not only test the performance of different MAB algorithms
but also design pull sequences to understand user preference dynamics
and test existing assumptions on user preferences. 
For example, one may design pull sequences
to study the correlation between rewards obtained at different time steps
and rewards obtained from pulling different but related arms. 
It is worth noting that the attention checks in our system
are also customizable, allowing for diverse attention checks for each recommendation domain.

We plan to open source our code and data. %
For more details about the experimental platform and MTurk-related implementation details,
we refer the readers to  Appendix~\ref{appendix:implementation-details} in the supplementary material.

\subsection{Recruitment and compensation}
\label{sec:recruitment}

To ensure the quality of our collected data,
we only allow MTurk workers to participate in the study
if they are U.S. residents
and have completed at least $500$ Human Intelligence Tasks (HITs) with an above 
$97\%$ HIT approval rate.
The anticipated (and actual) time to complete the study is 
less than $30$ minutes.
For participants who have successfully completed the study 
and answered the attention check questions correctly $70\%$ of time,
we have paid $\$7.5$ 
(the equivalent hourly salary is above $\$15/\text{hr}$).
In Appendix~\ref{appenidx:ui-attention},
we provide more details
on the participants' demographics and backgrounds.
Out of the $360$ participants who have successfully completed the study, 
$316$ passed the attention check 
(acceptance rate $87.8 \%$).
Our analyses are only conducted 
on the data collected from these participants.
Table~\ref{tbl:participants} shows 
the number of participants for each experimental setup.

\section{Evolving Preferences in $K$-armed bandits}\label{sec:evolving-preference}%

\begin{table*}
\centering
\begin{tabular}{@{}cccccccc@{}}
\toprule
                                           & Self-selected & UCB & TS & ETC & $\varepsilon$-Greedy & CYCLE & REPEAT \\ \midrule
\# of Participants                         & 74            & 40  & 44 & 39  & 41     & 40    & 38       \\
\bottomrule
\end{tabular}
\caption{
Number of participants for each algorithm. 
The description for Self-selected, UCB, TS, ETC and $\varepsilon$-Greedy are in Section~\ref{sec:mab}. 
CYCLE and REPEAT are defined in Section~\ref{sec:evolving-preference}.
}
\label{tbl:participants}
\end{table*}

As we have previously discussed, 
in $K$-armed bandits,
the reward distribution of each arm is assumed to be fixed over time~\citep{robbins1952some,slivkins2019introduction,lattimore2020bandit},
which implies that 
the mean reward of each arm remains the same. 
It is unclear whether such an assumption 
is reasonable in recommender system settings
where the reward distributions represent
human preferences. 
Our first aim is to test for the existence of evolving preference in the $K$-armed bandits setup.  
In other words, 
using randomized controlled trials, 
we want to %
test the following hypothesis:  
    \emph{
    In a $K$-armed bandit recommendation setting, the reward distribution of each arm (i.e., the user preference towards each item) 
    is not fixed over time.
    }

To answer this, 
we collect enjoyment scores 
for two fixed recommendation sequences,
where each sequence is of length $T=50$.
The sequence consists of recommendations
from $K=5$ genre of comics.
In other words, the total number of arms is $5$. 
The first sequence pulls the arms in a cyclic fashion which we denote by {CYCLE},
while the second sequence pulls each arm repeatedly for $m=T/K$ times which we denote by {REPEAT}: 
\begin{align*}
    &\text{CYCLE}: \; (\underbrace{12 \ldots K 12 \ldots K \ldots 12 \ldots K}_{12\ldots K \text{ for }m \text{ times}} ), \\
    &\text{REPEAT}: \; (\underbrace{22 \ldots 2}_{m \text{ times }} \underbrace{1 \ldots 1}_{m \text{ times}} 3 \ldots 3 \ldots K \ldots K).%
\end{align*}
We note that for both sequences, 
the number of arm pulls of each arm is the same ($m$ times). 
Since the order of the comics is fixed for each arm (e.g., pulling an arm for $m$ times will always result in the same sequence of $m$ comics from that genre), 
the set of comics recommended by the two sequences are the same.
The only difference between the two sequences is the order of the presented comics. 
Intuitively, if the mean reward of each arm is fixed over time 
(and does not depend on the past pulls of that arm), 
then the (empirical) mean reward of each arm should be very similar under the two pull sequences. 

In this work, we utilize a modification of the two-sample permutation test~\citep{fisher1936design} 
to deal with the different numbers of participants 
under the two recommendation sequences. 
We let $\pr{CYCLE}$ and $\pr{REPEAT}$ 
denote the set of participants assigned to the CYCLE and REPEAT recommendation sequence, respectively. 
For each participant $i \in \pr{CYCLE}$, %
we use $a_i(t)$ to denote the pulled arm  (the recommended comic genre) at time $t$
and $X_i(t)$ to denote the corresponding enjoyment score (the reward) that the participant has provided. 
Similarly, for each participant $j \in \pr{REPEAT}$, 
we use $a_j(t)$ and $Y_j(t)$ to denote the arm pulled at time $t$ and the enjoyment score collected from participant $j$ at time $t$. 
Using these notations, for each arm $k \in [K]$, 
we define the test statistic as follows:
\begin{align*}
    \tau_k &= \underbrace{\frac{1}{|\pr{CYCLE}|} \sum_{i \in \pr{CYCLE}} \left(\frac{1}{m} \sum_{t\in [T]: a_i(t) = k} X_i(t) \right)}_{M_k^\text{CYCLE}} \\
    &\; - \underbrace{\frac{1}{|\pr{REPEAT}|}   \sum_{j \in \pr{REPEAT}} \left( \frac{1}{m} \sum_{t\in [T]: a_j(t) = k} Y_j(t) \right)}_{M_k^\text{REPEAT}}.
\end{align*}
The test statistic $\tau_k$ is the difference between the mean reward (enjoyment score) $M_k^\cycle$ under the CYCLE recommendation sequence and  
the mean reward $M_k^\repeated$ under the REPEAT  
recommendation sequence for arm $k$. 
A non-zero $\tau_k$ suggests that the mean reward of arm $k$ is different under \cycle~and \repeated 
and that the reward distribution is evolving. 
The higher the absolute value of $\tau_k$ is, 
the bigger the difference between the two mean rewards is.
A positive test statistic value indicates that the participants prefer the arm under \cycle~over~\repeated~on average. 

To quantify the significance of the value of the test statistic, 
we use a two-sample permutation test to obtain the $p$-value of the test~\citep{fisher1936design}: 
First, we permute participants between $\pr{\cycle}$
and $\pr{\repeated}$ uniformly at random for $10,000$ times
and ensure that the size of each group remains the same after each permutation.
Then, we recompute the test statistic $\tau_k$ for each permutation to obtain a distribution of the test statistic.
Finally, we use the original value of our test statistic along with this distribution to determine the $p$-value. 

To report the $95\%$ confidence interval of the test statistic for each arm,
we use bootstrap and re-sample the data for $5,000$ times at the level of arm pulls. 
That is, for each arm pull at time $t$, 
we obtain its bootstrapped rewards under CYCLE and REPEAT 
by resampling from the actual rewards obtained by pulling that arm under CYCLE and REPEAT at time $t$, respectively. 
Given that we have conducted $5$ tests simultaneously (one for each arm),
in order to control the family-wise error rate,
we need to correct the level $\alpha_k$ $(k \in [K])$ for each test.
More formally, to ensure the probability that we falsely reject \emph{any} null hypothesis to be at most $\alpha$, 
for each test, the $p$-value of a test should be at most its corresponding corrected $\alpha_k$. 
We adopt the Holm's Sequential Bonferroni Procedure (details presented in Appendix~\ref{appendix:bonferroni}) 
to perform this correction~\citep{abdi2010holm}.

\begin{table*}
\centering
\begin{tabular}{@{}ccccccc@{}}
\toprule
       &   & family             & gag                 & political (conservative) & office              & political (liberal) \\ \midrule
\multirow{3}{*}{Overall} &
$\tau_k$ value & $0.290$              & $0.132$               & $0.445$                    & $0.047$               & $0.573$               \\
& $95\%$ CI     & $[0.042, 0.549]$ & $[-0.096, 0.371]$ & $[0.158, 0.744]$       & $[-0.196, 0.290]$ & $[0.301, 0.837]$  \\
& $p$-value   & $0.025^*$              & $0.275$                & $0.004^*$                    & $0.694$               & $<0.001^*$ %
\\ \midrule
\multirow{3}{*}{Heavy} &
$\tau_k$ value & $-0.394$              & $-0.556$               & $-0.694$                    & $-0.647$               & $-0.664$               \\
& $95\%$ CI     & $[-0.678, -0.114]$ & $[-0.864, -0.243]$ & $[-1.056, -0.325]$       & $[-0.944, -0.350]$ & $[-1.021, -0.302]$  \\
& $p$-value   & $0.007^*$              & $<0.001^*$                & $<0.001^*$                    & $<0.001^*$               & $<0.001^*$ %
\\ \midrule
\multirow{3}{*}{Light} &
$\tau_k$ value & $0.784$              & $0.635$               & $1.274$                    & $0.552$               & $1.475$               \\
& $95\%$ CI     & $[0.421, 1.150]$ & $[0.298, 0.977]$ & $[0.860, 1.681]$       & $[0.198, 0.905]$ & $[1.120, 1.827]$  \\
& $p$-value   & $<0.001^*$              & $<0.001^*$                & $<0.001^*$                    & $0.001^*$               & $<0.001^*$ %
\\ \bottomrule
\end{tabular}
\caption{
The difference between the mean reward under \cycle~and the mean reward under~\repeated~for each arm. 
All results are rounded to $3$ digits.
The $p$-values are obtained through permutation tests with $10,000$ permutations.
We use asterisk to indicate that the test is significant at the level $\alpha=0.1$. 
The $95\%$ confidence intervals are obtained using 
bootstrap with $5,000$ bootstrapped samples.
}\label{tbl:result-evolving-preference}
\end{table*}

Our results show that for three arms---family, political (conservative) and political (liberal)---the non-zero difference between the mean reward under the two recommendation sequences are significant at level $\alpha = 0.1$
(Table~\ref{tbl:result-evolving-preference}).
These findings confirm our research hypothesis that user preferences are not fixed (even in a short amount of time)
in a $K$-armed bandit recommendation setting. 
There may be many causes of the evolving reward distributions (preferences). 
One possibility, among many others, 
is that the reward of an arm depends on its past pulls.
In other words, people's preference towards an item depends on their past consumption of it. 
For example, existing marketing and psychology literature has suggested that people may experience hedonic decline upon repeated exposures to the same item~\citep{galak2018properties,baucells2007satiation}.
On the other hand, in music play-listing,
one may expect the expected reward of an arm (a genre) to increase due to the taste that the listener has developed for that genre of music~\citep{schafer2010makes}.
For a more comprehensive discussion on preference formation,
we refer the readers to~\citet{becker1996accounting}.

\begin{figure*}[h]
    \centering
    \includegraphics[width=\textwidth]{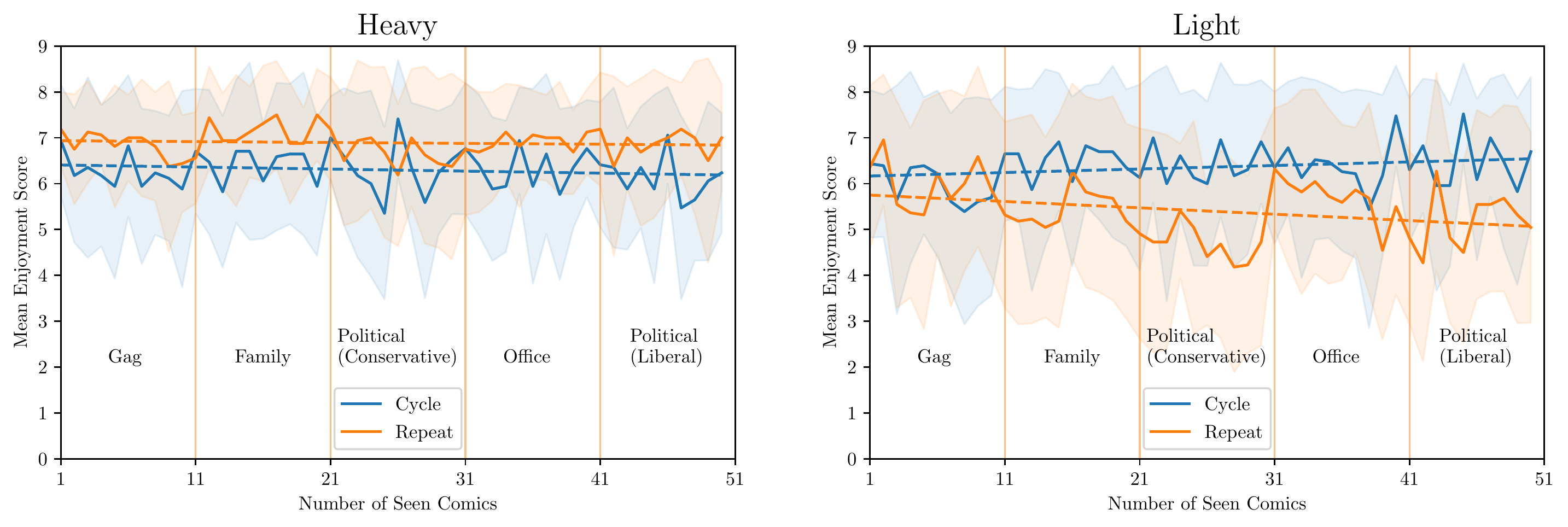}
    \caption{
    The reward at each time step of the CYCLE and REPEAT recommendation sequence averaged across heavy and light readers, respectively. {The error bars indicate one standard deviation from the mean.}
    For the REPEAT sequence, 
    we highlight when the arm switches using the vertical lines
    and add the arm name (comic genre) corresponding to the time period in between the switches using the black texts.
    The blue and orange dotted lines are fitted through the rewards collected under CYCLE and REPEAT, respectively.
    }
    \label{fig:stratified_mean_sequence}
\end{figure*}

\begin{figure*}[h]
    \centering
    \includegraphics[width=1\textwidth]{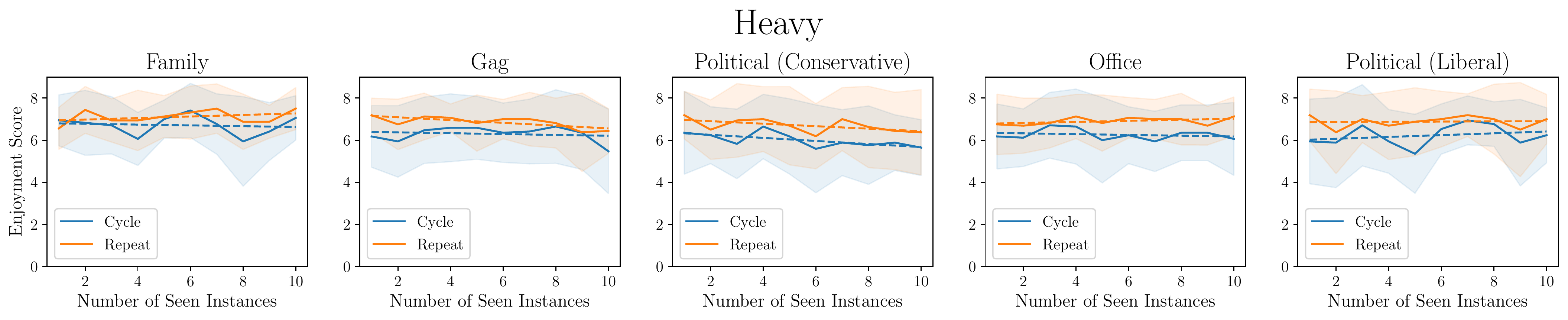}\vspace{1mm}
    \includegraphics[width=1\textwidth]{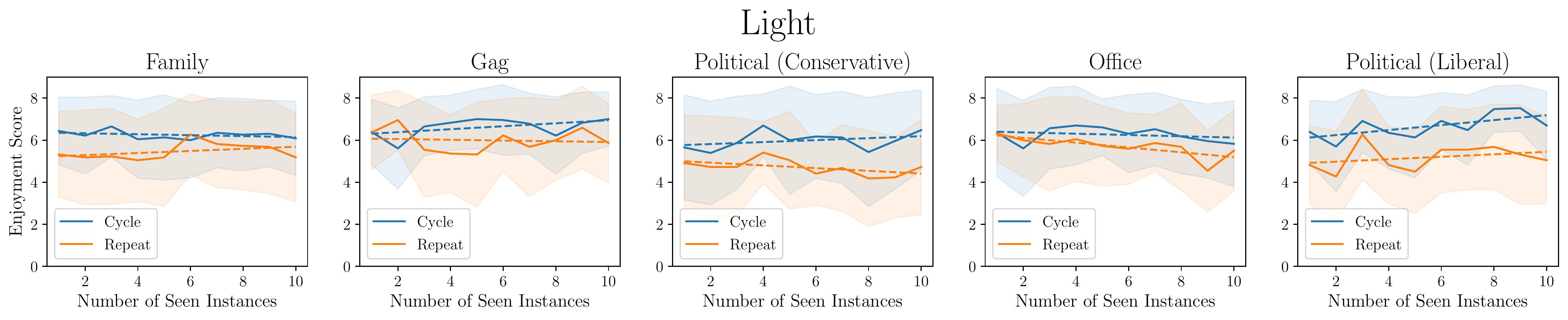}
    \caption{
    Each plot shows
    the reward collected for a particular arm at each arm pull  under the CYCLE and REPEAT recommendation sequences. {The error bars indicate one standard deviation from the mean.} The reward trajectories are averaged across heavy and light readers, respectively. 
    The blue and orange dotted lines are fitted through the reward trajectories for each arm under CYCLE and REPEAT, respectively.}\label{fig:stratified_mean_sequence_by_arm_heavy_readers}
\end{figure*}

Finally, 
to better understand the nature of our findings, 
we divide the participants 
into heavy comic readers who read comics daily 
and light comic readers who read comics at a lower frequency.
Among participants who are assigned the CYCLE sequence, 
there are  $17$ heavy readers and $23$ light readers. 
For REPEAT, there are $16$ heavy readers and $22$ light readers. 
We perform the same analysis as noted above for each of the two groups.
Similar to the overall findings,
among both heavy and light readers,
evolving preferences (evolving mean reward) have been observed  (Table~\ref{tbl:result-evolving-preference}),  
confirming our research hypothesis. 
Interestingly, we find that for each genre,
the heavy readers tend to 
prefer the recommendations from the REPEAT sequence over the CYCLE sequence.
On the contrary, for each genre,
light readers prefer recommendations from the CYCLE sequence over the REPEAT sequence.
As an initial step towards understanding this phenomenon, we present descriptive data analysis on how rewards (user preferences) evolve for heavy and light readers under the two recommendation sequences.
Similar to our results in Table~\ref{tbl:result-evolving-preference}, 
for light readers, at most time steps,
the reward trajectory of CYCLE  has a higher value than the reward trajectory of  REPEAT;
while for heavy readers, this is the opposite
(Figure~\ref{fig:stratified_mean_sequence}).
By looking at the reward trajectories (and the lines fitted through the reward trajectories) over the entire recommendation sequence (Figure~\ref{fig:stratified_mean_sequence}), we find additional trends:
for light readers, the differences between the rewards collected under CYCLE and REPEAT are increasing over time;
while for heavy readers, such differences are relatively stable.
This trend is also observed in 
the reward trajectory (and the line fitted through the reward trajectory) of each arm under CYCLE and REPEAT  (Figure~\ref{fig:stratified_mean_sequence_by_arm_heavy_readers}).
In particular, for light readers, among all arms (comic genres) except family, we find that the lines fitted through the rewards collected under CYCLE and REPEAT  become further apart as the number of arm pulls increases.
This distinction between heavy and light users may be due to various reasons. 
For example, light readers may prefer variety in their recommendations because they are exploring their interests 
or the light readers and heavy readers share different satiation rates.
On a related note,
a recent study has shown that the satiation rates of people may depend on their personality traits~\citep{galak2022identifying}.
Precisely understanding the causes of this  distinction between heavy and light readers is of future interest.


\section{Usage Example of the Experimental Framework}\label{sec:diff-algo}

As an illustration of the usage of our experimental framework, 
we compare the performance of different algorithms, 
in terms of both the cumulative rewards, 
and the users' own reflection on their interactions with these algorithms. 
Though we are in a %
simulated and simplified recommender system setup, 
we aim to provide some understanding on 
(i) whether people prefer to be recommended by an algorithm over deciding on their own, 
and (ii) whether more autonomy (choosing the next comic genre on their own) results in a more attentive and mindful experience.  
We note that, 
compared to our findings in Section~\ref{sec:evolving-preference} 
that are obtained through a rigorous hypothesis testing framework,
the results in this section are exploratory in nature and should not be interpreted as  definitive answers to the above questions.

\begin{table*}
\centering
\begin{tabular}{@{}cccccccc@{}}
\toprule
                                                         & Self-selected & UCB   & TS     & ETC    & $\varepsilon$-Greedy & CYCLE & REPEAT \\ \midrule
Cumulative reward  & $\begin{matrix} 319.36 \\ [205, 434] \end{matrix}$
                   & $\begin{matrix} 323.70 \\ [223, 424] \end{matrix}$
                   & $\begin{matrix} 312.37 \\ [204, 420] \end{matrix}$
                   & $\begin{matrix} 324.49 \\ [209, 440] \end{matrix}$
                   & $\begin{matrix} 307.59 \\ [205, 411] \end{matrix}$
                   & $\begin{matrix} 316.5 \\ [216, 417] \end{matrix}$ 
                   & $\begin{matrix} 301.63 \\ [201, 427] \end{matrix}$\\
Hindsight satisfaction         & $81.08 \%$          & $75.00 \%$ & $72.73 \%$  & $76.92 \%$  & $80.49 \%$          & $77.50 \%$ & $63.16 \%$  \\
Preference towards autonomy  & $71.62 \%$         & $50.00 \%$ & $68.18 \%$  & $58.97 \%$  & $58.54 \%$          & $62.50 \%$ & $65.79 \%$  \\ 
\bottomrule
\end{tabular}
\caption{Performances of each algorithm in terms of different enjoyment characterizations.
The first row gives the cumulative rewards for each algorithm, averaged over participants who have interacted with it.
We also report the $95\%$ confidence intervals for the cumulative rewards. 
The hindsight-satisfaction row shows the percentage of participants who believe the sequence of comics they have read captures their preference well. 
The last row provides the percentage of participants who prefer to select comics to read on their own in hindsight.
}
\label{tbl:performance}
\end{table*}

\subsection{Enjoyment}\label{sec:enjoyment}
We first compare different algorithms (Self-selected, UCB, TS, ETC, $\varepsilon$-Greedy, CYCLE and REPEAT)
in terms of the participants' enjoyment. 
More specifically, we want to compare  
the participants 
who are provided recommended comics to read 
with the ones who choose on their own. 
In general, enjoyment is hard to measure~\citep{payne1999measuring}.
To this end, 
we look at participants' enjoyment 
and preference towards these algorithms 
through the following three aspects: 
\begin{itemize}
    \item Cumulative rewards: This metric is closely related to the notion of ``regret'' that is commonly used to compare the performance of bandit algorithms~\citep{lattimore2020bandit}.
    More formally, for any participant $i$, the cumulative reward of an algorithm that interacts with participant $i$ is given by $\sum_{t=1}^T R_i(t)$ where $R_i(t)$ is the reward provided by participant $i$ at time $t$. 
    In Table~\ref{tbl:performance}, for each algorithm, 
    we show their cumulative rewards averaged over participants who have interacted with the algorithm.
    \item Hindsight satisfaction: 
    After the participants interact with the algorithm, 
    in the post-study survey, we ask them the following question: ``Do you feel that the sequence of recommendations\footnote{For Self-selected participants, this should be interpreted as comics they have read/self-selected.} captured your preferences well?''
    Participants' answers to this question provide their hindsight reflections towards how well the algorithm  performed and whether they are satisfied with it. 
    The second row of Table~\ref{tbl:performance} shows the percentage of participants who believe the sequence of comics they read has captured their preference well. 
    \item Preference towards autonomy: 
    In addition to the previous two metrics, 
    we have explicitly asked the participants to indicate whether they prefer to choose comics to read on their own. In the post-study survey, we ask: ``Do you prefer being recommended comics to read or selecting comics to read on your own?'' 
    The third row of Table~\ref{tbl:performance} provides the percentage of participants who prefer to select comics on their own for each algorithm. 
\end{itemize}

In our collected data, 
the participants who have given more autonomy (the Self-selected participants) 
prefer more autonomy in hindsight, compared to other participants. 
Though Self-selected does not have
the highest mean cumulative reward, 
it has 
the highest percent 
of participants 
who 
believe 
that the comics they read 
have captured their preferences well in hindsight 
 (Table~\ref{tbl:performance}). 
This misalignment between mean cumulative reward 
and hindsight satisfaction also shows in other algorithms 
(i.e., higher mean cumulative reward may not indicate higher hindsight satisfaction), 
including $\varepsilon$-Greedy and ETC. 
On the other hand, UCB performs well %
in terms of the mean cumulative reward,
while having the lowest percentage of 
participants wanting to have more autonomy. 
We provide additional analyses on comparing these algorithms in terms of participant's enjoyment in Appendix~\ref{appendix:additional-analysis}.

\begin{table*}
\centering
\begin{tabular}{@{}cccccccc@{}}
\toprule
                                               & Self-selected & UCB   & TS    & ETC   & $\varepsilon$-Greedy & CYCLE & REPEAT \\ \midrule
Reading Memory                                 & $91.89\%$         & $90.00\%$ & $92.42\%$ & $93.16\%$ & $90.24\%$ & $90.00\%$ & $90.35\%$  \\
Rating Memory & $71.17\%$         & $70.00\%$ & $70.45\%$ & $76.92\%$ & $69.92\%$ & $69.16\%$ & $57.89\%$  \\ \bottomrule
\end{tabular}
\caption{Average correctness 
of the two types of memory questions for each algorithm.
}
\label{tbl:memory}
\end{table*}

\subsection{Attentiveness}\label{sec:memory}
We want to 
understand whether participants 
who are asked to choose on their own
have a more mindful experience,
in which the participants are more attentive 
to what they have read through~\citep{epstein1999mindful}.
There is no consensus on defining and measuring mindfulness of an experience~\citep{grossman2008measuring}. 
Some of the existing research uses self-reported mindfulness
as a measurement~\citep{greco2011assessing}. 
In our case, 
we look at how attentive the participants are 
to their own experience, 
though the lens of memory---for each participant, 
in the post-study survey,
we ask them two types of memory questions:
\begin{itemize}
    \item Reading memory: 
    In the post-study survey, 
    we present the participants three randomly selected comics and ask them to indicate whether they have read the comics before. %
    This question aims to measure the participants' memory of \emph{what they have read}. 
    The first row of Table~\ref{tbl:memory} shows the average correctness percentage for the reading memory questions for each algorithm.  
    \item Rating memory: 
    We also look at the participants' memory on \emph{how they have liked} certain comics. 
    To this end, in the post-study survey, we present three randomly selected comics among the ones the participants have read and ask them to indicate whether they have rated the comic with a score of five or above. 
    The second row of Table~\ref{tbl:memory} shows the average correctness percentage. 
\end{itemize}
We note that these questions differ from 
the attention check questions after each comic
and are not directly related to the compensation
that the participants get. 
However, 
the high correctness percentage shown in Table~\ref{tbl:memory} %
on the reading memory questions
suggest that the participants have attempted to 
answer the questions from their memory
instead of providing random answers.

In general, the participants have performed 
much better on the reading memory questions
than the rating memory questions, 
suggesting that they may be more aware of what they have consumed than how they have liked them. 
In addition, all participants 
perform similarly in terms of the reading memory correctness percentage 
with those whose algorithm is ETC or Self-selected 
performing slightly better. 
Though it is hard to say which algorithm provides the most attentive experience to the participants 
and whether Self-selected participants have a more mindful experience, 
it is relatively clear that participants whose algorithm is REPEAT 
perform the worst in terms of rating memory. 
Our data does not provide strong evidence on believing that 
more autonomy results in more attentive experience, 
but may suggest that less enjoyable experience (e.g., for  participants whose algorithm are REPEAT)
correlates to less attentiveness. 
We provide additional analyses on comparing these algorithms in terms of participant's attentiveness in Appendix~\ref{appendix:additional-analysis}.

\section{
{Conclusions and Future Work}}
\label{sec:discussion}

Our work provides a general experimental framework and toolkit
to understand assumptions on human preferences in a bandit setup.
It also allows one to 
perform field tests of different bandit algorithms
that interact with humans.
Using these tools, 
we build 
a publicly available 
dataset that contains 
trajectories of recommendations 
and ratings of them given by multiple users  
in a bandit setting. 
The collected data has been used to 
check the validity of a core assumption on human preference in MABs literature---that the reward distributions (corresponding to user preferences) are fixed.
We show that even in a short time period like our bandit recommendation setup, such dynamical preference exists. 
Further, our observations on the difference between the light and heavy user in their preference dynamics suggest the need of having a more granular understanding of human preferences. 
As an illustrative usage of our experimental framework, 
we have explored the study participants' 
preferences towards selecting content to read on their own
and being recommended content to read. 
As we have discussed above, these findings are exploratory in nature with the goal of showcasing different usages of our experimental framework; thus, they should not be interpreted as definitive answers.
In our exploratory analysis, we observe that an algorithm achieving the highest cumulative reward does not necessarily imply that it will achieve the highest hindsight satisfaction. 

At a higher level, our work fits in the broader picture of understanding the validity of assumptions that machine learning systems (and in our case, recommender systems) rely on.
Although all machine learning models are built upon some simplifying assumptions of the world, some of these assumptions  oversimplify the world to the extent that they lose the core characterizations of the problem we are interested in. 
In our case, the assumptions we want to understand are centered around user preferences. 
We have identified that assumptions on the temporal stability of preferences used in traditional MABs literature are  oversimplifications. 
The balance between identifying simplifications that are helpful for building learning systems and avoiding oversimplifications that discard core characteristics of the problem is difficult. 
As put by the famous statistician George Box, ``all models are wrong, but some are useful'' \citep{box1979all}.
Our goal  for developing the experimental toolkit and conducting the human subjects study is to provide ways for identifying assumptions on user preferences that are useful for developing MABs algorithms in recommender systems.
Below we discuss the limitations 
and future directions of our work.

\subsection{Limitations}

We discuss several limitations of our study. 
First,
the sizes of the mean reward differences of each arm reported in Section~\ref{sec:evolving-preference} 
are within $1.5$ point on the $9$-point Likert scale, 
which may be considered small. 
This is due to many reasons. 
For one, the enjoyment score (the reward)
provided by each participant is subjective
and may have high variance due to this subjectivity. 
Though a difference may be considered to be strong in a within-subjective study,
it may be thought of as small in a between-subject study (the type of study in our case).
However, we note that our results 
obtained using the permutation test
show that reward distributions are indeed not fixed over time
for multiple arms in the $K$-armed bandit recommendation setup. 
Second, each arm represents a comic series.
Though we have selected the comics from each series
in terms of their quality (the number of likes that the comics receive), 
there may still be heterogeneity among the selected comics belonging to the same series,
and it is up to discussion on whether one should consider 
these comics to belong to the same arm. 
Thirdly, many quantities (e.g., enjoyment/satisfaction, mindfulness/attentiveness) we want to measure 
are less well-defined. 
Our way of measuring them is from a particular angle,
and may not be widely applicable. 
Fourthly, the study domain is chosen to be comics due to reasons including the short consumption time of a comic and the study requires the participants to read $50$ comics. Although all of our experiments are completed within $30$ minutes, it is uncommon in real-world settings for people to read $50$ comics at a time and thus may introduce uncontrolled boredom effects.
Fifthly, in our experiments, we compare the user preferences towards each arm using two fixed sequences. One may also utilize a purely randomized sequence as a baseline to better understand user preferences. 
Finally, due to resource constraints (e.g., funding limits for conducting the study and computational limits on the number of  simultaneous experiment instances we can host on our server), we recruited $360$  participants (with $316$ of them passing the attention checks) for our study. Compared to industry-scale experiments, the number of participants in our study is on the lower end.
{It is also worth mentioning that our implementations of the bandit algorithms ensure that 
the comic recommendations only depend on the user's own history, which is not a common practice for recommender systems on existing  platforms. In practice, one may utilize other users' interaction histories to warm start these bandit algorithms.}
Though there are these limitations,
we would like to emphasize that our work 
makes a substantive step towards 
understanding the applicability of traditional
assumptions on user preferences in MABs literature.

\subsection{Future Work}

There are multiple future directions for our work. 
Our findings on the existence of evolving preferences 
in a $K$-armed bandits recommendation setting  
suggest that in order to study the decision-theoretic nature
of recommender systems using the MAB framework,
one must account for such preference dynamics. 
The need for learning algorithms (oftentimes reinforcement learning algorithms)
that deal with the impact of 
recommendations on user preferences 
have also been proposed in recent works~\citep{zheng2018drn, ie2019recsim, shani2005mdp, mcinerney2018explore, chen2019top, mehrotra2020bandit, chen2021user, chen2021values, wang2022surrogate, chen2022off}.
An important building block for this line of research
is to have better modeling of human preference dynamics. 
Our experimental framework and toolkit 
can provide more grounding and accessibility for research on it.
As an example, 
our observation that heavy and light comic readers have different preference dynamics can be further investigated using our experimental framework,
advancing our understanding on evolving preferences in a more granular way.
More broadly, our toolkit can be used for:
(i) collecting data using fixed or randomized recommendation sequences
or bandit algorithms to identify and estimate preference dynamics; 
and (ii) conducting field tests 
of bandit algorithms designed to address evolving  preferences.

Our exploratory data analysis on 
the performance of different algorithms 
suggests that 
the human interactants of the bandit algorithms
may care about other aspects of their experience 
in addition to cumulative rewards.
For example, Self-selected has a higher percentage of satisfied participants in hindsight compared to UCB, though UCB has a higher average cumulative reward.  
This suggests that besides traditional performance metrics
used to analyze and develop these bandit algorithms, 
we should consider a broader set of objectives and metrics
when studying these problems. 
On a related note,
given that we want our algorithm to account for evolving preferences, 
when regret (the difference between the expected cumulative reward obtained by a proposed policy 
and an oracle) is used as the performance metric, 
the oracle should be chosen to be adaptive instead of the best-fixed arm considered in many MABs (including contextual bandits) literature.

\bibliographystyle{ACM-Reference-Format}
\bibliography{preference-dataset}

\appendix
\section{Algorithms}\label{appendix:algorithms}

We provide additional details 
on algorithms presented in Section~\ref{sec:mab}. 
Recall that $R(t)$ denote the reward obtained at time $t$
and $a(t)$ denote the arm pulled at time $t$.
Denote $\mathcal{T}_{k,t} = \{t' \in [t-1]: a(t') = k\}$. 
{As we discussed in our main paper, the reward $R(t)$ obtained by pulling arm $a(t)$ is given by the enjoyment score the user provides after reading the recommended comic. 
Thus, for each arm (i.e., each comic genre), the reward distribution corresponds to user's enjoyment score distribution (or in other words, user's preference) towards that genre. 
}
\begin{itemize}
    \item UCB: 
    For $t \in \{1, \ldots, K\}$, 
    we have $a(t) = t$. 
    For $t \in \{K+1, \ldots, T\}$, we have 
    $$a(t) \in \arg\max_{k \in [K]} \frac{1}{|\mathcal{T}_{k,t}|}\sum_{t \in \mathcal{T}_{k,t}} R(t) + \sqrt{\frac{2 \log t}{|\mathcal{T}_{k,t}|}}.$$
    \item TS: For each arm $k \in [K]$, their prior distribution is set to be a Dirichlet distribution with parameters being $ (1, 1, 1, 1, 1, 1, 1, 1, 1)$, 
    given that the rewards are categorical with $9$ values.
    At time $t$, for each arm $k$, we obtain a virtual reward 
    $\Tilde{R}(k,t)$ by sampling from the corresponding Dirichlet distribution first and then use that sample to sample from a multinomial distribution. Then, 
    $a(t) \in \arg\max_{k \in [K]} \Tilde{R}(k,t)$. 
    Upon a reward $R(t)$ is received by pulling arm $k$, 
    the parameter of arm $k$'s Dirichlet distribution is updated:
    the $R(t)$-th entry of the parameter is increased by $1$.
    {We have chosen the Dirichlet and multinomial distribution as the prior distribution and likelihood function because of their conjugacy, which ensures that the posterior distribution is tractable. This choice of the prior distribution and likelihood function has not utilized the fact that the reward is ordinal. One can potentially use a likelihood function that captures the ordinal structure of the rewards, but it may result in difficulties in obtaining the posterior distribution since there may not be a conjugate prior for this new likelihood function.}
    \item ETC: For $t \in \{1, \ldots, \lfloor 0.5 \cdot (T^{2/3})\rfloor \}$, $a(t) = t \text{ mod } K$.
    For $t \in \{\lceil 0.5 \cdot (T^{2/3})\rceil, 
    \ldots, T\}$, 
     $$a(t) \in \arg\max \frac{1}{|\mathcal{T}_{k,t}|}\sum_{t \in \mathcal{T}_{k,t}} R(t).$$
    \item $\varepsilon$-Greedy: 
    For $t \in \{1, \ldots, K\}$, $a(t) = t$.
    For $t \in \{K+1, \ldots, T\}$, 
    with probability $0.9$, 
    $$a(t) \in \arg\max \frac{1}{|\mathcal{T}_{k,t}|}\sum_{t \in \mathcal{T}_{k,t}} R(t).$$
    With probability $0.1$, 
    $a(t)$ is randomly selected from the $K$ arms.
\end{itemize}

\section{Holm’s Sequential Bonferroni Procedure}
\label{appendix:bonferroni}
{
When testing $K$ hypotheses (in our case, we have one hypothesis for each arm),
we adopt Holm’s Sequential Bonferroni Procedure to control the family-wise error rate. 
To ensure that the probability of falsely rejecting \emph{any} null hypothesis to be at most $\alpha$, we perform the following procedure:
Suppose that we are given $m$ sorted $p$-values $p_1, \ldots, p_m$ from the lowest to the highest for hypothesis $H_1, \ldots, H_m$.
For $i \in [m]$, if 
$p_i < \alpha/(m+1-i)$, 
then reject $H_i$ and move on to the next hypothesis; otherwise, exit the process (and we cannot reject the rest of the hypotheses).
As an illustration, 
in Table~\ref{tbl:result-evolving-preference},
since we are testing $5$ hypothesis (one for each arm) at the same time and we have set the overall $\alpha$ level to be $0.1$,
to reject all null hypotheses,
we require the lowest to the highest $p$-values to be no bigger than 
 $0.02, 0.04, 0.06, 0.08, 0.1$.
}
In other words, setting $\alpha$ at level $0.1$ suggests that the probability of falsely rejecting \textit{any} null hypothesis is at most $0.1$.
Under Bonferroni’s correction, when rejecting the null hypothesis with the lowest $p$-value, the corrected alpha level is $\alpha/K = 0.02$.

\section{Additional analyses for Section~\ref{sec:diff-algo}}
\label{appendix:additional-analysis}
{We present
additional analyses comparing the performance of different MABs algorithms in terms of user enjoyment and attentiveness.
As discussed in the main paper, these analyses are meant to showcase potential use cases of our experimental framework instead of claiming that certain algorithms outperform others.  
For user enjoyment, we focus on users' hindsight satisfaction (Section~\ref{sec:enjoyment}) for  different algorithms and use the two-sample permutation test to check whether the algorithms' performances differ much from Self-selected's. 
More specifically, for each algorithm (UCB, TS, ETC, $\epsilon$-Greedy, CYCLE and REPEAT), we compare its  hindsight satisfaction rate with the rate for Self-selected. 
The test statistic is defined as follows: 
\begin{align*}
    \beta_\text{Algo} &= \text{Hindsight Satisfaction under Algo} \\
    &- \text{Hindsight Satisfaction under Self-selected}.
\end{align*}
Under the null hypothesis, $\beta_\text{Algo} = 0$, which indicates that the algorithm and Self-selected share similar performances in terms of the hindsight satisfaction rate.
The results shown in Table~\ref{tbl:hindsight_satisfaction_comparision} suggest that 
in our experiment, although on average Self-selected performs better than other algorithms in terms of hindsight satisfaction, we cannot reject the null hypotheses that these algorithms' hindsight satisfaction is similar to Self-selected's in general.
The only null hypothesis we can reject is the one for REPEAT. 
That is, we can claim that in terms of hindsight satisfaction rate, Self-selected outperforms REPEAT. 
In general, collecting more data may help us increase the power of our tests.
In terms of attentiveness, 
we study the rating memory of the participants (Section~\ref{sec:memory}) under different algorithms. 
Similarly, we adopt the two-sample permutation test to check whether other algorithms' performances differ much from that of Self-selected, in terms of the the percentage of questions that the participants answer correctly regarding the ratings they provided for comics they read. 
The specific test statistic we used is as follows:
\begin{align*}
    \omega_\text{Algo} &= \text{Correctness Percentage under Algo} \\
    &- \text{Correctness Percentage under Self-selected}.
\end{align*}
The null hypothesis states that the correctness percentage on these rating-memory-related questions should not differ much between other algorithms and Self-selected. 
Using the data from our experiments, we cannot reject this null hypothesis. 
However, as demonstrated here, our experimental framework equips researchers with tools to answer these types of questions. 
}

\begin{table*}[h]
\centering
\begin{tabular}{@{}ccccccc@{}}
\toprule
                                 & UCB   & TS    & ETC  & Greedy & CYCLE  & REPEAT \\ \midrule
 $\beta_\text{Algo}$ value   & -6.08\%  & -8.35\% & -4.16\% & -0.59\% &  -3.59\% & -17.9\% \\
 $95\%$ CI  & [-9.53\%, 22.64\%]  & [-7.06\%, 24.26\%] &  [-11.50\%, 20.76\%] &  [-14.30\%, 16.02\%] & [-11.82\%, 19.40\%] & [0.57\%, 35.28\%] \\
                         $p$-value & $0.15$  & $0.095$  & $0.22$ & $0.36$  &  $0.24$  & $<0.01^*$ \\ \bottomrule 
\end{tabular}
\caption{
{The difference between the hindsight satisfaction rate  under Algo (UCB, TS, Greedy, CYCLE and REPEAT) and the hindsight satisfaction rate under Self-selected. 
All results are rounded to $2$ digits.
The $p$-values are obtained through permutation tests with $10,000$ permutations.
We use asterisk to indicate that the test is significant at the level $\alpha=0.1$.} 
}\label{tbl:hindsight_satisfaction_comparision}
\end{table*}

\begin{table*}[h]
\begin{tabular}{@{}ccccccc@{}}
\toprule
                                & UCB    & TS     & ETC     & Greedy & CYCLE  & REPEAT  \\ \midrule
$\omega_\text{Algo}$ value   & 1.17\% & 0.72\% & -5.75\% & 1.25\% & 2.00\% & 13.28\% \\
 $95\%$ CI  & [-12.70\%, 15.43\%]  & [-13.62\%, 15.34\%] & [-18.31\%, 7.21\%] & [-11.72\%, 14.34\%] &  [-10.77\%, 14.71\%] & [-1.42\%, 28.02\%] \\
                         $p$-value & $0.42$   & $0.46$   & $0.18$    & $0.44$   & $0.37$   & $0.04$   \\ \bottomrule
\end{tabular}
\caption{
{The difference between the correctness percentage on rating-memory questions under Algo (UCB, TS, Greedy, CYCLE and REPEAT) and the correctness percentage under Self-selected. 
All results are rounded to $2$ digits.
The $p$-values are obtained through permutation tests with $10,000$ permutations.
}} 
\label{tbl:rating_memory_comparision}
\end{table*}

\section{Background Survey}
\label{appenidx:ui-attention}
\begin{figure}[h!]
    \centering
    \begin{subfigure}{0.45\textwidth}
        \includegraphics[width=0.95\textwidth]{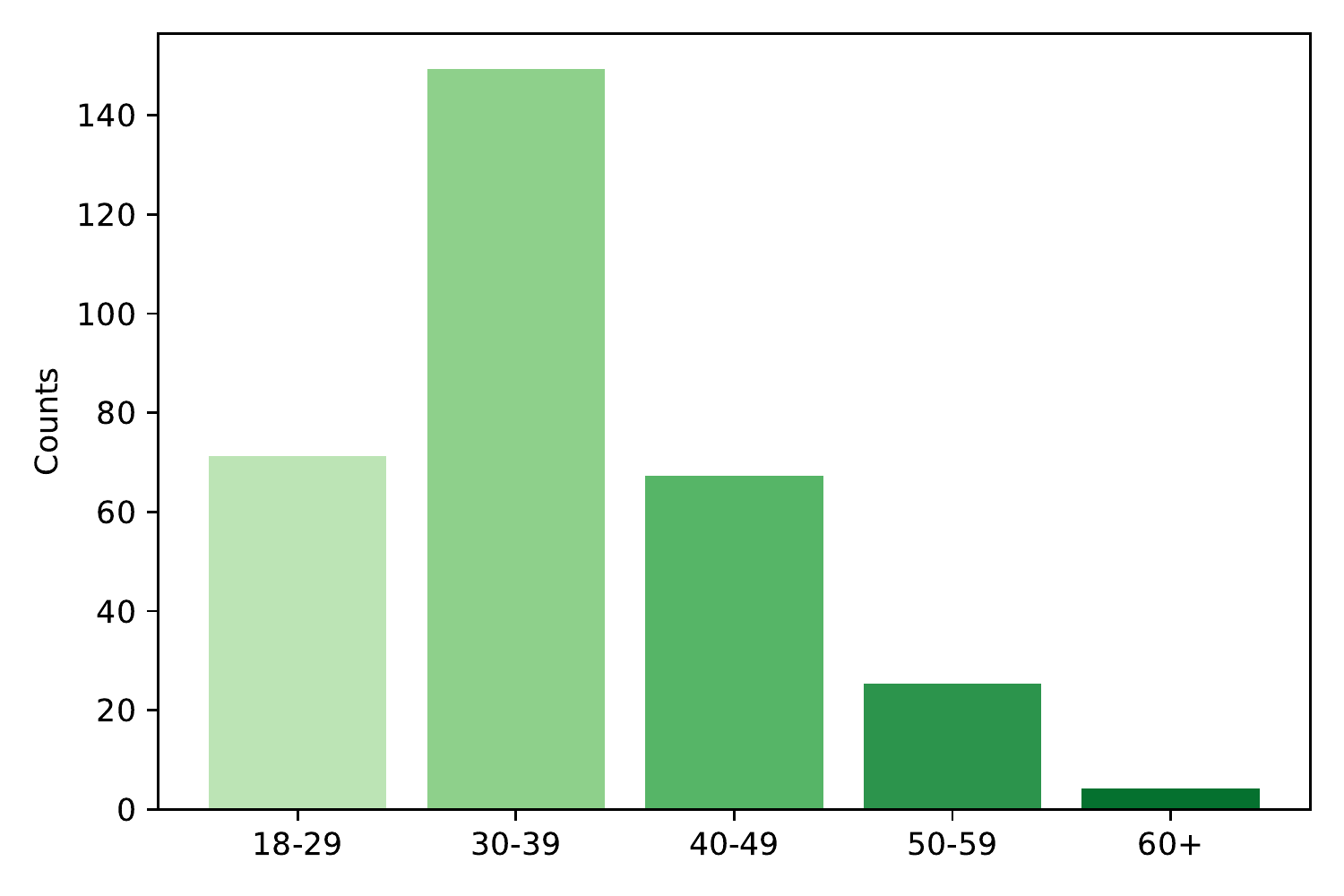}
        \caption{Participant age distribution.}
    \end{subfigure}
    \begin{subfigure}{0.45\textwidth}
        \includegraphics[width=0.95\textwidth]{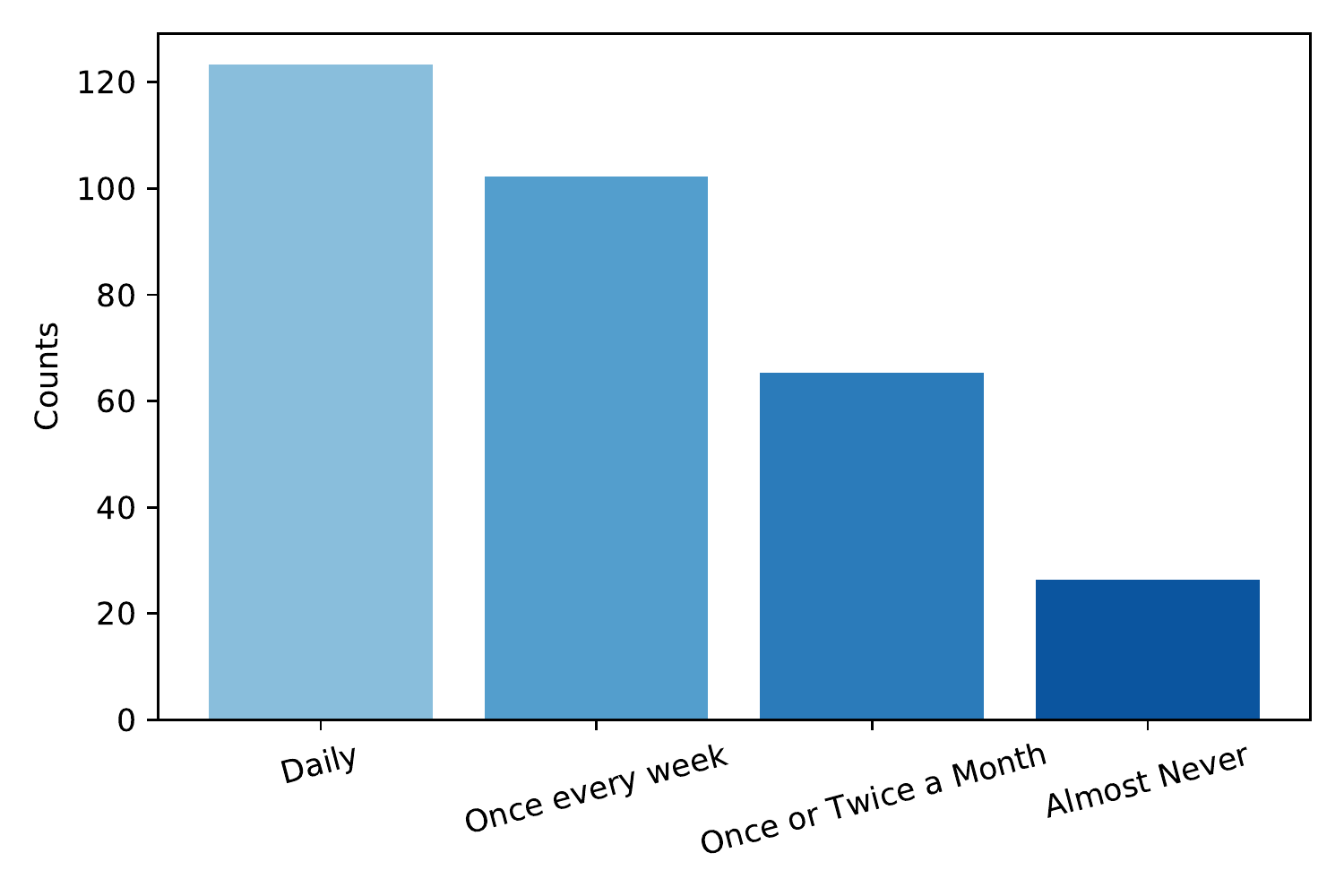}
        \caption{How often participants read comics.}
    \end{subfigure}\vspace{1em}
    \begin{subfigure}{0.45\textwidth}
        \includegraphics[width=0.95\textwidth]{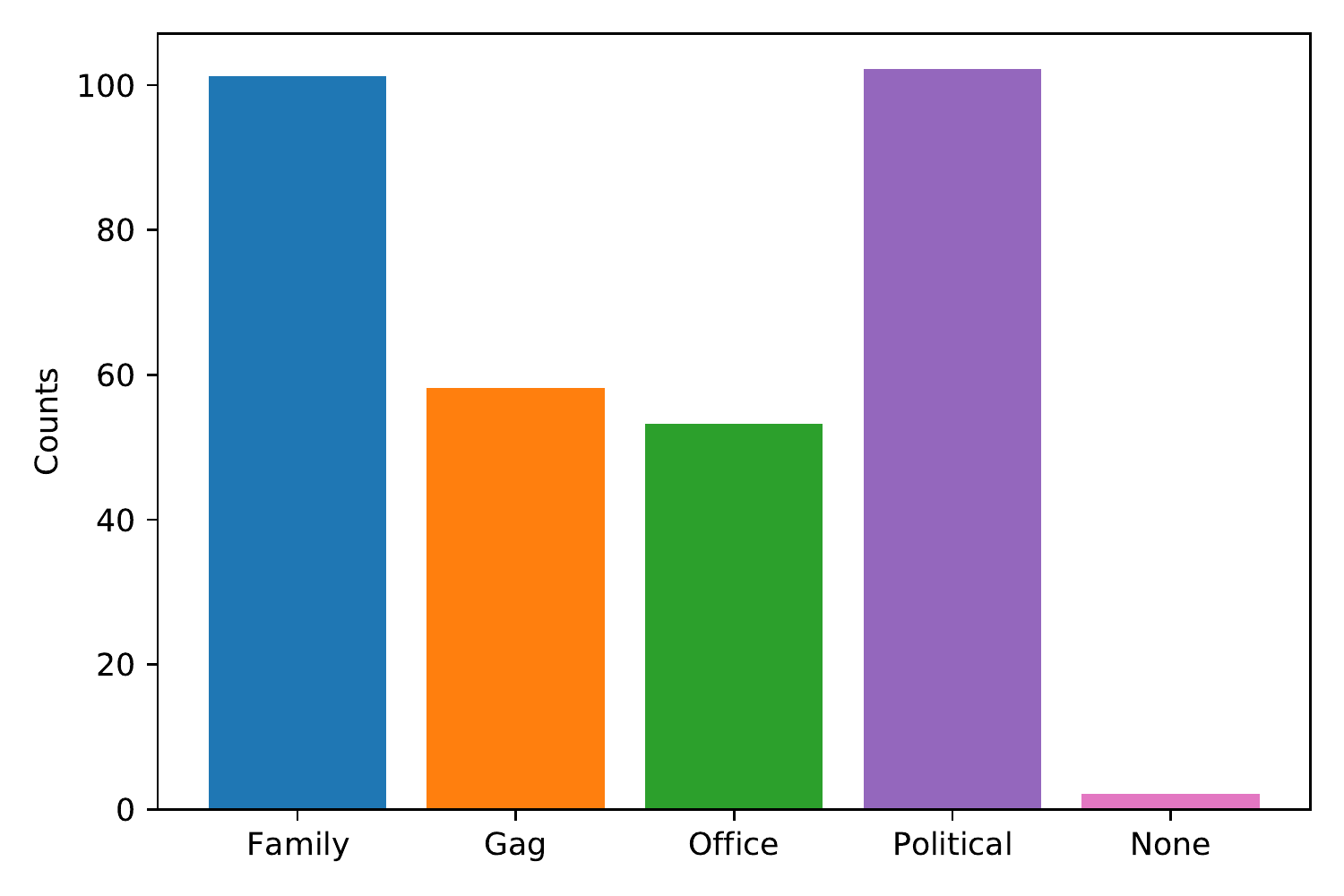}
        \caption{Participants' most familiar category of comic.}
    \end{subfigure}
    \begin{subfigure}{0.45\textwidth}
        \includegraphics[width=0.95\textwidth]{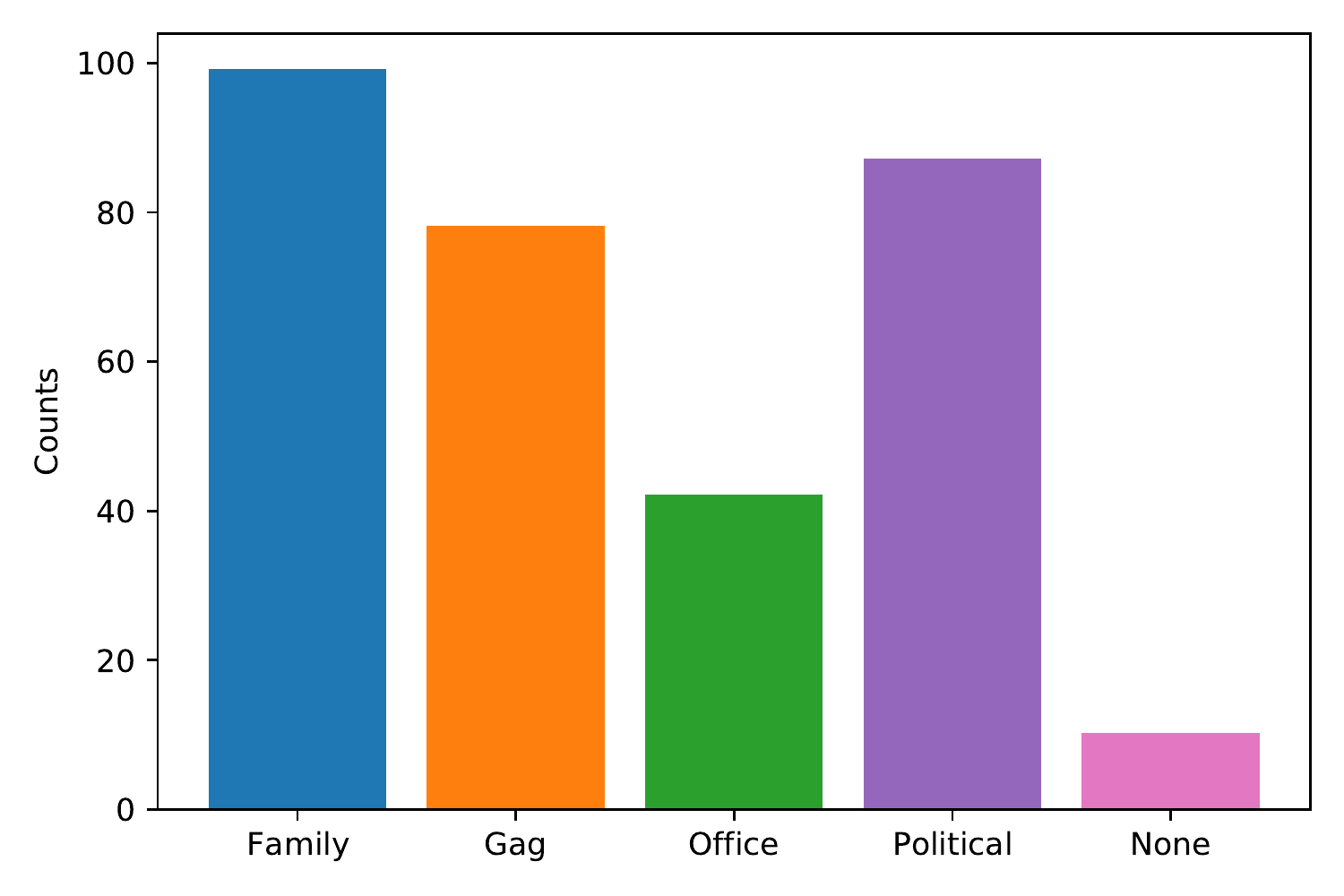}
        \caption{Participants' most liked category of comic.}
    \end{subfigure}
    \caption{Histograms of demographic information collected from the background survey.}
    \label{fig:demographicsl}
\end{figure}

We asked the study participants the following four questions:
\begin{itemize}
    \item What is your age?
    \item How often do you read comic strips (e.g. newspaper comics, Sunday comics, ...)? Please pick an option that best describes you. 
    \item Out of the following genres of comic strips, which genre do you find to be the most familiar?
    \item Out of the below genres of comic strips, which genre do you like the most?
\end{itemize}
The provided data are plotted as histograms in Figure~\ref{fig:demographicsl}.

\section{Implementation Details}
\label{appendix:implementation-details}

\subsection{Platform implementation} 
As described in Section~\ref{sec:framework}, our platform consists of two main components: the participant-facing web interface and server backend. In this section, we provide more in-depth implementation details about both platform components.

\paragraph{Web interface.} The participant-facing web interface is written using standard web development tools (HTML/CSS/Javascript) as well as the Flask web framework.
When participants are performing the comic rating portion of the study, their responses are submitted in the background.
Once the response is successfully recorded by the server, the images, rating slider bar, and button orderings are updated in-place so that users are not distracted by full page refreshes after each comic.
The web interface also has a experimenter-defined variable to enforce a minimum time that must elapse between submissions.
This ensures that participants spend an adequate amount of time on each comic; there is also a server-side mechanism to handle duplicate submissions if the button is clicked multiple times in quick succession.
Once the comic reading portion is complete, the web interface will automatically transition to the post-study survey after users have read the end-of-study message.

\paragraph{Server.} We use the Flask web framework to implement the server and MongoDB for the database backend.
The MAB algorithms are implemented in Python with a common interface that defines \texttt{get\_arm()} and \texttt{update\_arm()} functions, which can easily be extended to include other MAB algorithms that share this interface.
The user accounts used by participants, the comic rating responses, post-study survey responses, and attention check answers are each stored in a separate database.
All experiment configurations are prepared prior to launching the study.
Experiments are configured using JSON files that enable both local and global control over experiment parameters, such as the number of comics to read, or the post-study survey and attention check questions asked.
Participants can only access their assigned experiment, using credentials that are given when they register on the website.
As participants are completing tasks in the study, each submitted task response is recorded in the database.
The server can use the stored data to reload the session where it left off if the browser is closed or refreshed.
This state includes that required for the algorithms, which are all executed on the server and require up-to-date parameters at each timestep of the study. 
Lastly, we host the server on Amazon Elastic Compute Cloud (EC2) using a \texttt{t2.xlarge} instance with 4 vCPUs and 16GB of memory.

\subsection{MTurk implementation} The study is split into two separate tasks on the MTurk platform.
Participants must sign up for both tasks in order to proceed with the study. 
Once they have completed the initial background survey with a valid completion code, they must register their MTurk ID on our website.
If their MTurk ID is found among the completed Qualtrics survey entries (retrieved through the MTurk query API), they are then assigned an account with a pre-loaded configuration.
We also assign a unique survey code to each participant in the event to ensure that all parts of the study are properly completed.
Submissions that contain reused survey codes from other participants are considered invalid, and therefore not included in the final dataset.

\onecolumn

\begin{figure*}
    \centering
    \begin{subfigure}{0.8\textwidth}
        \includegraphics[width=\textwidth]{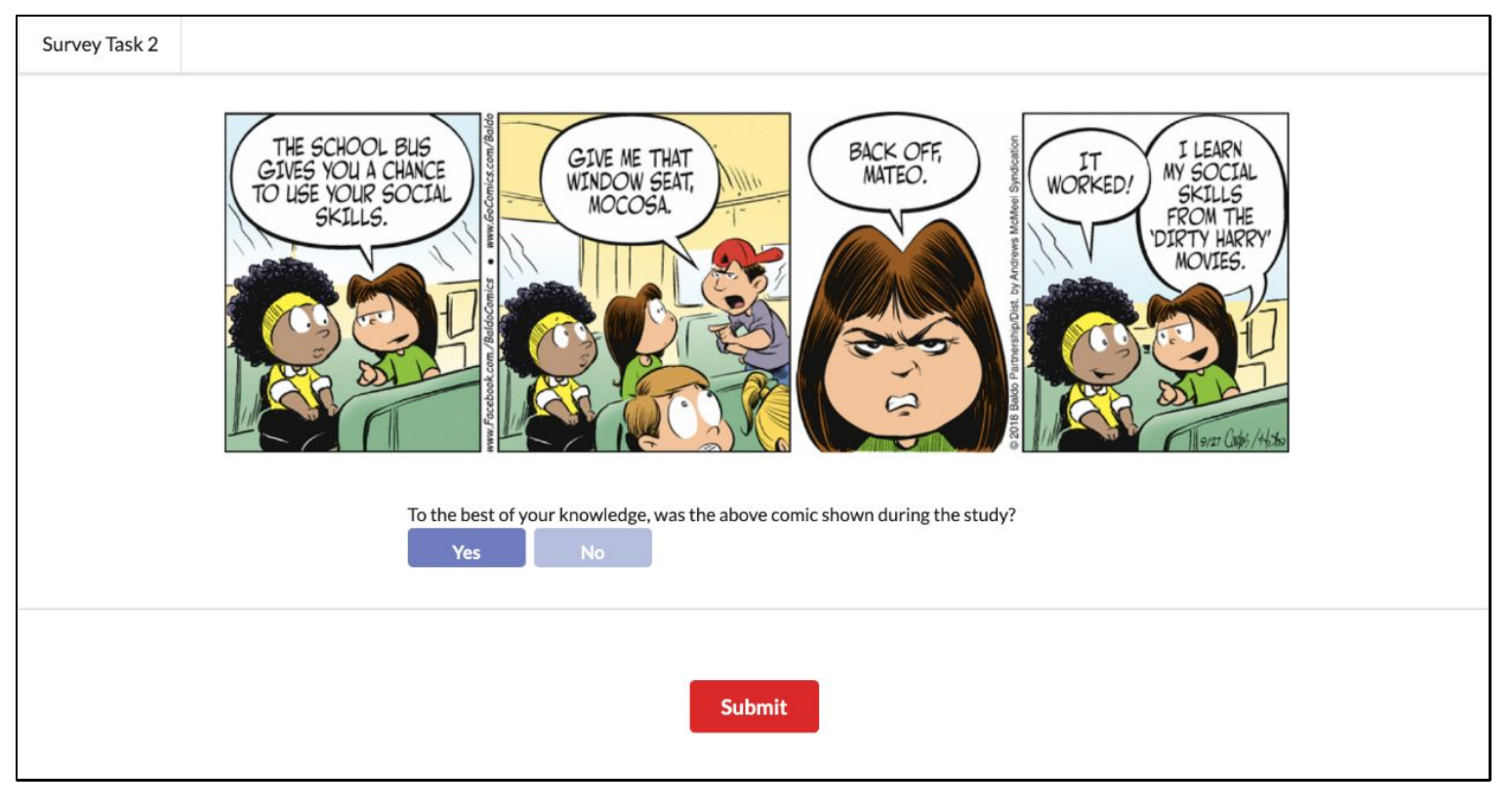}
        \caption{Testing whether the participant remembers if the given comic was in the study.}
    \end{subfigure}
    \begin{subfigure}{0.8\textwidth}
        \includegraphics[width=\textwidth]{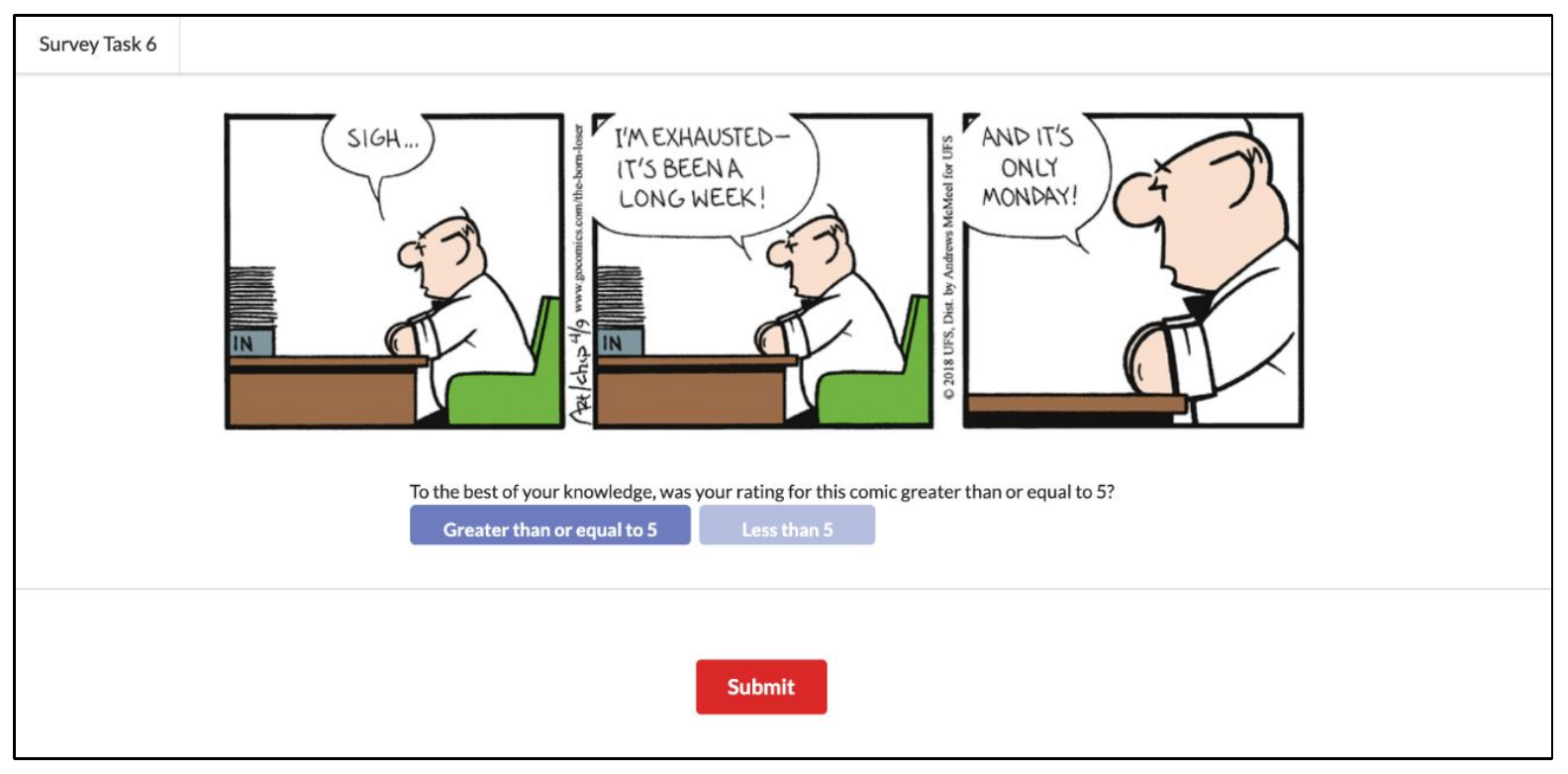}
        \caption{Testing whether the participant remembers rating the given comic positively in the study.}
    \end{subfigure}
    \begin{subfigure}{0.8\textwidth}
        \includegraphics[width=\textwidth]{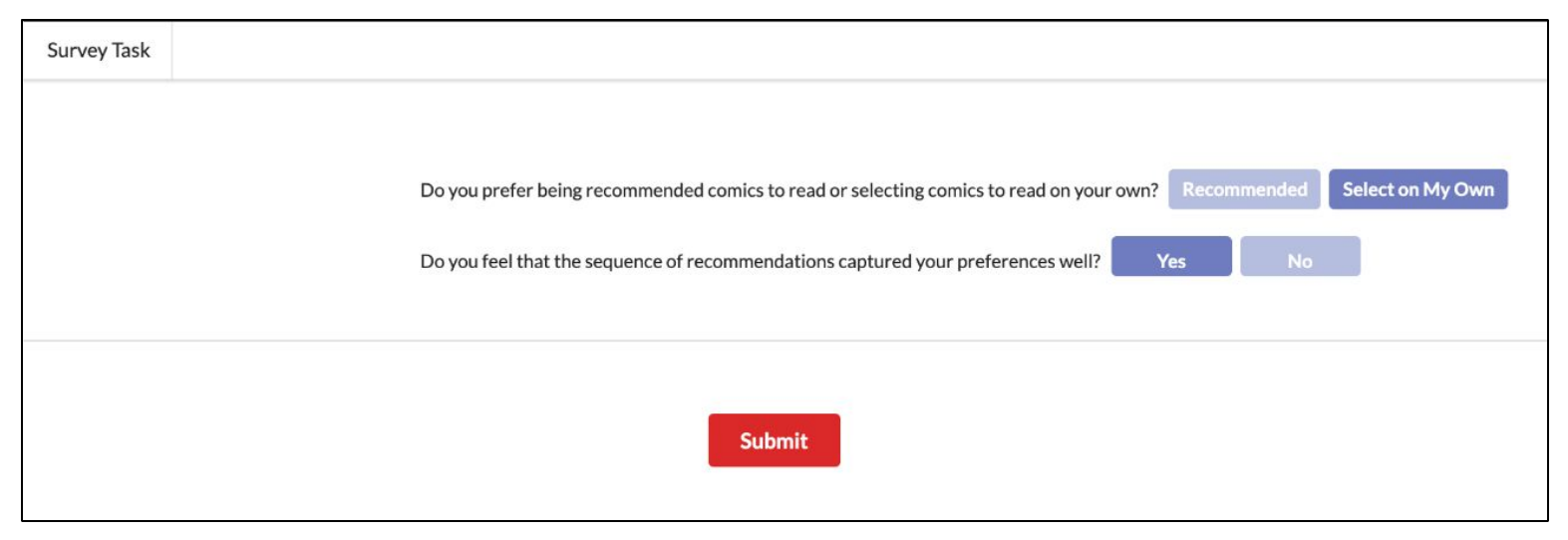}
        \caption{Final questions asked in the post-study survey.}
    \end{subfigure}
    \caption{An example of questions contained in the post-study survey.}
    \label{fig:post-study-survey}
\end{figure*}

\end{document}